\documentclass[aps,prd,reprint,preprintnumbers,superscriptaddress,nofootinbib,longbibliography,floatfix]{revtex4-2}
\pdfoutput=1
\usepackage{rotating}
\usepackage{array}
\usepackage{amsmath}
\usepackage[normalem]{ulem}
\usepackage{slashed}
\usepackage{booktabs}
\usepackage[pdftex,table]{xcolor}
\usepackage{units}
\usepackage{xfrac}
\usepackage{mathtools}
\usepackage{empheq}
\usepackage[]{units}
\usepackage{multirow}
\usepackage{amssymb}
\usepackage{url}
\usepackage{comment}
\usepackage{physics}
\usepackage{color,soul}
\usepackage{bbm}
\usepackage[caption=false]{subfig}
\usepackage{adjustbox}
\usepackage{hyperref}

\usepackage{hyperref}
\hypersetup{
  colorlinks=true,
  citecolor=blue,
  linkcolor=blue,
  urlcolor=blue
}
\newcommand{\inv}{^{-1}}

\newcommand{\qt}{{\widetilde{q}}}

\newcommand{\bins}{\textrm{bins}}
\newcommand{\N}{\mathcal{N}}

\usepackage{color}
\definecolor{darkred}{rgb}{1.0,0.1,0.1}
\definecolor{darkgreen}{rgb}{0.1,0.7,0.1}
\definecolor{darkblue}{rgb}{0.1,0.1,1.0}

\DeclareRobustCommand{\Sec}[1]{Sec.~\ref{sec:#1}}

\DeclareRobustCommand{\App}[1]{App.~\ref{app:#1}}

\DeclareRobustCommand{\Tab}[1]{Table~\ref{tab:#1}}

\DeclareRobustCommand{\Fig}[1]{Fig.~\ref{fig:#1}}

\DeclareRobustCommand{\Eq}[1]{Eq.~(\ref{eq:#1})}

\DeclareRobustCommand{\Reff}[1]{Ref.~\cite{#1}}
\DeclareRobustCommand{\Reffs}[1]{Refs.~\cite{#1}}
\allowdisplaybreaks

\begin{document}

\title{Moment Unfolding}

\author{Krish Desai}
\email{krish.desai@berkeley.edu}
\affiliation{Department of Physics, University of California, Berkeley, CA 94720, USA}
\affiliation{Physics Division, Lawrence Berkeley National Laboratory, Berkeley, CA 94720, USA}

\author{Benjamin Nachman}
\email{bpnachman@lbl.gov}
\affiliation{Physics Division, Lawrence Berkeley National Laboratory, Berkeley, CA 94720, USA}
\affiliation{Berkeley Institute for Data Science, University of California, Berkeley, CA 94720, USA}

\author{Jesse Thaler}
\email{jthaler@mit.edu}
\affiliation{Center for Theoretical Physics, Massachusetts Institute of Technology, Cambridge, MA 02139, USA}
\affiliation{The NSF AI Institute for Artificial Intelligence and Fundamental Interactions}

\preprint{MIT-CTP 5727}

\begin{abstract}
Deconvolving (``unfolding'') detector distortions is a critical step in the comparison of cross section measurements with theoretical predictions in particle and nuclear physics.
However, most existing approaches require histogram binning while many theoretical predictions are at the level of statistical moments.
We develop a new approach to directly unfold distribution moments as a function of another observable without having to first discretize the data.
Our Moment Unfolding technique uses machine learning and is inspired by Boltzmann weight factors and Generative Adversarial Networks (GANs).
We demonstrate the performance of this approach using jet substructure measurements in collider physics.
With this illustrative example, we find that our Moment Unfolding protocol is more precise than bin-based approaches and is as or more precise than completely unbinned methods.
\end{abstract}

\maketitle

{
\tableofcontents
}

\section{Introduction}
\label{sec:intro}

Studying the dependence of physical observables on various quantities like energy scale offers a rich probe into the complex scaling dynamics of fundamental physical theories.
In many cases, it is advantageous to summarize a probability distribution through a small number of statistical moments, which makes visualization and interpretation more tractable and lends itself to more precise theoretical predictions.
For example, the spectra of many quark and gluon jet observables cannot be computed from first principles in perturbative quantum chromodynamics (QCD), but the energy dependence of their moments can be precisely predicted from factorization and Dokshitzer–Gribov–Lipatov–Altarelli–Parisi (DGLAP) evolution~\cite{Altarelli:1977zs,Dokshitzer:1977sg,Gribov:427157}.
Additionally, one of the most precise extractions of the QCD coupling constant comes from comparing measured moments~\cite{AMY:1989feg,TASSO:1990cdg,MovillaFernandez:1997fr,OPAL:1997asf,DELPHI:1999vbd,L3:2000shd,DELPHI:2003yqh,DELPHI:2004omy,L3:2004cdh,ALEPH:2003obs,OPAL:2004wof,Pahl:2009zwz} to theoretical calculations~\cite{Abbate:2012jh}.

\textit{Unfolding}, also known as \textit{deconvolution}, is the process of correcting detector distortions in experimental data.
This is necessary for the accurate comparison of data between experiments, and with theoretical predictions.
Typically, entire spectra are unfolded and then moments are computed afterward.
In order to capture the dependence of an observable $Z$'s moments on another quantity $Y$, these two features must be simultaneously unfolded.
Current unfolding approaches discretize the $(Z,Y)$ support, and then the two-dimensional histogram is unfolded such that the moments of $Z$ can be computed in bins of $Y$.
This binning procedure introduces discretization artifacts that hinder comparisons between measurements and theory, and between data from different experiments.

One possibility to improve the extraction of moments from data is to unfold without binning.
A number of unbinned unfolding techniques have been proposed, including many based on machine learning~\cite{Zech:2003rsn,Lindemann:1995ut,Datta:2018mwd,bunse2018unification,2019ASPC..521..394R,Andreassen:2019cjw,Bellagente:2019uyp,1800956,Vandegar:2020yvw,Andreassen:2021zzk,Howard:2021pos,Backes:2022vmn,Arratia:2022wny,Chan:2023tbf,Shmakov:2023kjj,Alghamdi:2023emm,Diefenbacher:2023wec,Pan:2024rfh} (see \Reffs{Arratia:2021otl,Huetsch:2024quz} for overviews).
In terms of experimental applications, the OmniFold method~\cite{Andreassen:2019cjw,Andreassen:2021zzk} has recently been applied to studies of hadronic final states with data from H1~\cite{H1:2021wkz,H1prelim-22-031,H1:2023fzk,Nachman:2022ltz}, LHCb~\cite{LHCb:2022rky}, CMS~\cite{Komiske:2022vxg,CMS-PAS-SMP-23-008}, STAR~\cite{Song:2023sxb}, and ATLAS~\cite{ATLAS:2024xxl}.
By construction, these unbinned approaches do not introduce binning artifacts. 
Nevertheless, because they offer a generic solution to unfolding entire spectra, unbinned methods may compromise precision for any particular aspect of the spectrum, such as a small set of moments.
Furthermore, existing unbinned methods are also mostly iterative~\cite{Andreassen:2019cjw,Andreassen:2021zzk,Backes:2022vmn,Pan:2024rfh}, which increases their computational complexity.

In this paper, we introduce a dedicated machine learning-based unfolding method to directly unfold moments of observable distributions.
Our Moment Unfolding technique is motivated by the Boltzmann distribution from statistical mechanics and uses a structure that is similar to Generative Adversarial Networks (GANs)~\cite{Goodfellow:2014upx}.
In particular, we learn a reweighting function at particle level whose form is determined by a Boltzmann weight factor so that its parameters can be identified with the observable moments.
This function is optimized by requiring that the reweighted simulation at detector level is as similar as possible to the target data (determined by a discriminator), similar to the two-level GAN setups in \Reffs{Chan:2023tbf,Chan:2023ume,Bierlich:2023zzd}.
Like OmniFold~\cite{Andreassen:2019cjw,Andreassen:2021zzk}, our approach is based on reweighting, but it is fundamentally different because it is not iterative.
We restrict our attention here to a small number of moments of a single observable.
In principle, this approach could be extended to multiple observables and even full distributions, which we leave for future studies.

The remainder of this paper is organized as follows.
We briefly review the statistics of moments in \Sec{moments} and how these quantities are can be measured using binned or unbinned approaches.
Our Moment Unfolding protocol is introduced in \Sec{momentunfolding}.
In \Sec{case}, we provide numerical case studies, first on a Gaussian toy example and then on a realistic particle physics study involving jet substructure observables.
The paper ends with conclusions and outlook in \Sec{conclusions}.

\section{The Statistics of Moment Measurements}
\label{sec:moments}

As a reminder, the $k^\text{th}$ moment of a probability density $p_Z(z)$ is calculated formally by taking an integral over the weighted density:\footnote{Upper-case letters represent random variables and lower-case letters represent realizations of those random variables.} 
\begin{align}
\label{eq:moment}
\langle Z^k\rangle \equiv \int_{-\infty}^\infty z^k \,p_Z(z) \,\dd z\,.
\end{align}
When studying the dependence of the $k^\text{th}$ moment of $Z$ on another observable $Y$, the probability density $p_Z(z)$ is replaced with the conditional probability density $p_{Z|Y}(z|y)$.
Throughout this paper, we restrict our attention to the case that $Z$ and $Y$ are one-dimensional observables.

\subsection{Biases from Binning}

To estimate the quantity in \Eq{moment}, one often uses a histogram approximation of $p_Z(z)$:
\begin{align}
\label{eq:binned}
\langle Z^k\rangle\approx \langle Z^k\rangle_\text{bin}\equiv \frac{1}{N}\sum_{i=1}^{n_\text{bins}} N_i\, z^k_{\text{bin},i}\,,
\end{align}
where $N_i$ is the number of counts in bin $i$, $N=\sum_i N_i$ is the total number of counts, and $z_{\text{bin},i}$ is the center of bin $i$.
In the limit that $N, n_\text{bins}\rightarrow\infty$ with equally spaced bins, $\langle Z^k\rangle_\text{bin}\rightarrow \langle Z^k\rangle$.%
\footnote{Often non-uniform bin spacing is used to accommodate non-linear detector resolutions.  The statement in the text is true more generally when the maximum bin width goes to zero.}

To see that binning generically leads to biases, one can rewrite \Eq{moment} as:
\begin{align}
\langle Z^k\rangle &= \lim_{n_{\bins}\to\infty}\sum_{i=1}^{n_\bins} \int_{z_i}^{z_{i+1}}  z^k \,p_Z(z)\, \dd z\\
&\equiv \lim_{n_{\bins}\to\infty} \sum_{i=1}^{n_\text{bins}} p_i\, \langle Z^k\rangle_i\,, 
\end{align}
where $\langle Z^k\rangle_i$ is the moment of $z$ in bin $i$ and $p_i$ is the fraction of $p_Z(z)$ that falls in bin $i$.
Therefore, in the limit that $N\rightarrow \infty$, but $n_\text{bins}$ is finite, the bias due to the binning is:
\begin{align}
\label{eq:bias}
\langle Z^k\rangle- \langle Z^k\rangle_\text{bin}=\sum_{i=1}^{n_\text{bins}} \left(\langle Z^k\rangle_i-z_{\text{bin},i}^k\right)\,.
\end{align}
This equation emphasizes that, if instead of using the bin centers as in \Eq{binned}, one were to use the $k^\text{th}$ moment per bin, then the binning bias could be removed.
The histogramming tool YODA~\cite{Buckley:2023xqh} keeps track of first and second moments within bins for precisely this reason.

For spectra that are monotonically increasing or decreasing, \Eq{bias} predicts the sign of the bias.
For spectra that have one or more maxima, it is not possible to even know, in general, if there is a bias and if so, what is the sign of the bias.

\subsection{Unfolding Binned Measurements}

In an experimental context, before computing $\langle Z^k\rangle_\text{bin}$ in \Eq{binned}, it is necessary to estimate $N_i$.
A variety of regularized matrix inversion approaches have been proposed, which use a response matrix $\mathbf{R}$ to relate the counts at detector level to the counts at particle level.
In particular, the folding equation can be written as $\mathbf{x}=\mathbf{R}\,\mathbf{z}$, where $\mathbf{x}$ and $\mathbf{z}$ are vectors with the detector-level and particle-level counts, respectively, and $\textbf{R}$ is the response matrix.
The elements of the response matrix are
\begin{align}
R_{ij}=\Pr(\text{measure in bin $i$} | \text{truth is bin $j$})\,,
\end{align}
where $\Pr(\cdot)$ indicates probability of the argument.
Note that in general, $\mathbf{R}$ need not be a square matrix, which is one reason why simple matrix inversion is not typically effective for unfolding.

The most common approaches to inferring $\mathbf{z}$ from $\mathbf{x}$ include Iterative Bayesian Unfolding (IBU)~\cite{DAgostini:1994fjx} (also known as Richardson-Lucy deconvolution~\cite{Richardson:72,1974AJ.....79..745L}), Singular Value Decomposition (SVD)~\cite{Hocker:1995kb}, and TUnfold~\cite{Schmitt:2012kp}.
Reviews on unfolding methods can be found in \Reffs{Cowan:2002in,Blobel:2203257,doi:10.1002/9783527653416.ch6,Balasubramanian:2019itp}.
The method we will use as a baseline for the case studies in \Sec{case}  is IBU, which proceeds iteratively:
\begin{align}\nonumber
z_j^{(t)}&=\sum_i \text{Pr}^{(t-1)}(\text{truth is $j$}|\text{measure $i$})\,\Pr(\text{measure $i$})\\
&=\sum_i\frac{R_{ij}\,z_j^{(t-1)}}{\sum_m R_{im} \, z_m^{(t-1)}}\times x_i\,,
\label{IBU}
\end{align}
where $z^{(0)}$ is a prior distribution (often taken to be the particle-level simulation) and $t$ is the iteration number.

After unfolding, a number of classical approaches have been proposed to correct for the bias in \Eq{bias}.
Perhaps the most common is to apply a multiplicative correction to the binned unfolded data:
\begin{align}
\langle Z^k\rangle_\text{meas.}=\langle X^k\rangle_\text{bin, data}\times \frac{\langle Z^k\rangle_\text{MC truth}}{\langle X^k\rangle_\text{bin, MC reco.}}\,,
\end{align}
where $\langle Z^k\rangle_\text{MC truth}$ is the moment computed in simulation without binning.
A challenge with this approach is that it does not make use of any local information in $Z$, since all values that enter in the above equation are summed over bins.
To solve this, one could apply a correction per $Z$ bin:
\begin{align}
\label{eq:bincorrection}
\langle Z^k\rangle_\text{meas.}=\frac{1}{N}\sum_{i=1}^{n_\text{bins}} N_i \, \langle Z^k\rangle_\text{MC truth,\,$i$}\,,
\end{align}
where $\langle Z^k\rangle_\text{MC truth,\,$i$}$ is the mean value of $X$ in the $i^\text{th}$ bin from simulation.
In this case, one relies on the prior density within a given bin of $Z$, but if the prior is not too different from nature, the resulting bias will be suppressed.
It may be possible to further improve this by using data to unfold also the values of $\langle Z^k\rangle_{i}$.

\subsection{Unbinned Unfolding Methods}
\label{sec:omnifold}

One way to completely avoid the bias in \Eq{bias} is to unfold without binning in the first place.
There are a number of unbinned unfolding methods, but to our knowledge, the only one applied to data so far is OmniFold~\cite{Andreassen:2019cjw,Andreassen:2021zzk}, which generalizes IBU.
It uses neural network classifiers to iteratively reweight the particle- and detector-level Monte Carlo events, respectively.
The final product of OmniFold is a weighting function $\nu(z)$ so that the unfolded expectation value of any observable computable from $Z$ is given by:
\begin{align}
\label{eq:observe_from_omnifold}
\langle\mathcal{O}\rangle=\sum_{z\in\mathrm{Gen}} \nu(z)\, \mathcal{O}(z)\,,
\end{align}
where the sum runs over synthetic particle-level events from a Monte Carlo generator.
Usually, $\mathcal{O}$ represents the counts in a given bin of a histogram from a differential cross section measurement.
However, $\mathcal{O}$ could also be the $k^\text{th}$ moment of $Z$ directly.
A key benefit of OmniFold is that all unfolded expectation values are derived simultaneously from a single reweighting function.

Because of its generality, OmniFold may not be as precise for any particular observable and moment.
In principle, OmniFold is capable of unfolding all observables and all moments simultaneously and studies have shown that adding more features can improve the precision on a given observable~\cite{Andreassen:2019cjw}.
However, the same studies have also shown that adding more information can reduce precision.
This may be due to cases where the gains from new information covariate with the detector response are outweighed by the additional regularization needed to fit higher-dimensional data.
Detailed studies of this bias-variance trade off would be interesting to explore in the future.

A computational challenge with OmniFold and most other methods that actively mitigate prior dependence~\cite{Andreassen:2019cjw,Andreassen:2021zzk,Backes:2022vmn,Pan:2024rfh} is that they are iterative.
In practice, this means that unfolding may require training tens of neural networks (one for each step of each iteration), which can easily reach thousands when ensembling is added into the workflow to achieve stability.  
\section{Moment Unfolding}
\label{sec:momentunfolding}
\begin{figure*}[t]
    \centering
    \includegraphics[width=0.90\textwidth]{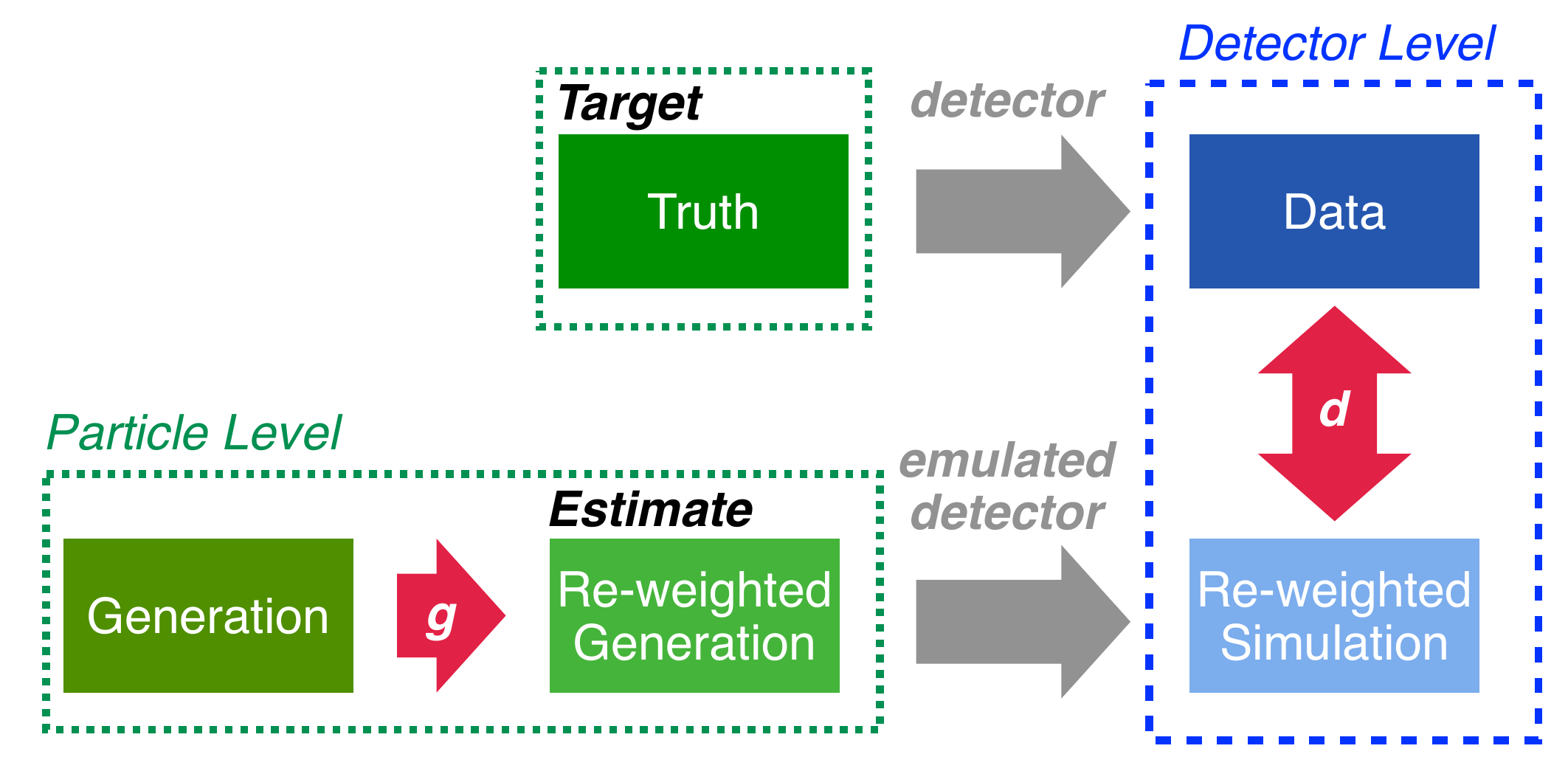}
    \caption{A schematic diagram of the training setup for Moment Unfolding.  Like a GAN, $g$ is the generator and $d$ is the discriminator, but now $g$ is simply a reweighting factor given by \Eq{generator}.  The reweighted simulation dataset inherits its weight from the matching generation dataset.  The detector emulations are only run once, since a new simulated dataset is created via importance weights and not by changing the features themselves.}
    \label{fig:schematic}
\end{figure*}

Motivated by the above challenges, we introduce  Moment Unfolding, which can directly learn moments without first unfolding the entire spectrum.
Moment Unfolding is an unbinned, non-iterative, reweighting-based method to unfold the statistical moments of observables, inspired by Boltzmann's approach to construct the Maxwell--Boltzmann distribution~\cite{e17041971}.

For the following discussion, we use the nomenclature of \Reff{Andreassen:2019cjw}, where unfolding involves four datasets: truth, data, generation, and simulation.  
Each synthetic collision event comes as a pair $(Z,X)$, for $Z\in\mathbb{R}^{N_{\mathrm{Gen}}}$ and $X\in\mathbb{R}^{N_{\mathrm{Sim}}}$, where $Z$ is the pre-detector version of the event (``generation'') and $X$ is the post-detector observation of the event (``simulation'').
In experimental data, we only have access to the detector-level version (``data''), so we use the simulation to infer the underlying pre-detector distribution (``truth'').

\subsection{Leveraging Boltzmann Weights}

Moment Unfolding uses a weight function $g(z)$, similar to OmniFold's $\nu(z)$ in \Sec{omnifold}.
Instead of determining the weight functions in an iterative fashion, though, Moment Unfolding uses a fixed functional form: 
    \begin{equation}
        \label{eq:generator}
        g(z) = \frac1{P}\exp\Bigg[-\sum_{a = 1}^n\beta_a\,z^a\Bigg],
    \end{equation}
where $n$ is the number of moments to be simultaneously unfolded, $\beta_a$ are parameters to be determined, and $P$ is a normalization constant, similar to the partition function from statistical mechanics.
When we want to unfold moments conditional on another observable $y$, the parameters $\beta_a$ are replaced with functions $\beta_a(y)$, as discussed more in \Sec{momentum_dependence}.
The exponential form of \Eq{generator} is inspired by the $e^{-\beta E}$ Boltzmann factor from statistical physics, whose derivation is reviewed in \App{boltzmann}.

To better understand this choice, recall that the Maxwell--Boltzmann distribution is the one that maximizes the entropy of an ensemble while holding mean energy constant.
This logic can be extended beyond means to arbitrary constraints, yielding the maximum entropy probability distribution~\cite{e17041971}.
In this language, \Eq{generator} optimizes the relative entropy of the reweighted truth-level distribution with respect to the $Z$ prior, while holding the first $n$ moments fixed to some value.

As described in more detail in \Sec{ml_implementation}, we determine the values of $\beta_a$ by maximizing the Maximum Likelihood Classifier loss~\cite{DAgnolo:2018cun,DAgnolo:2019vbw,Nachman:2021yvi}
between the reweighted detector-level simulation and experimental data.%
\footnote{See \Reff{Botev2011TheGC} for related discussions using the Binary Cross Entropy loss.}
Crucially, the learned values of $\beta_a$ are \emph{not} the learned moments themselves.
Rather, analogously to \Eq{observe_from_omnifold}, the moments are given by 
\begin{equation}
\label{eq:moment_calc_from_weights}
\langle Z^k \rangle_{\rm Moment\,Unfolding} = \sum_{z\in \mathrm{Gen.}} g(z)\,z^k\,,
\end{equation}
where the sums run over synthetic particle-level events, and the normalization $P$ is determined numerically:
\begin{equation}
\label{eq:partition_function}
P = \sum_{z\in \mathrm{Gen.}} \exp\Bigg[-\sum_a\beta_a z^a\Bigg].
\end{equation}
In this way, the extracted $k^\text{th}$ moment depends on all $n$ $\beta_a$ values.%
\footnote{This distinction is why we use $k$ to index the measured moments but $a$ to index the learned parameters.}

There is some arbitrariness in the choice of $g(z)$, since many weight functions with $n$ free parameters can in principle be used to match $n$ moments of a distribution.
The advantage of our choice of $g(z)$ is that, for the training procedure described below, Moment Unfolding provably converges to the truth moments under certain conditions, as described in \App{analytic}.

The hyperparameter $n$ sets the degree of the polynomial in the exponent, i.e.\ the number trainable weights in the generator and consequently the number of moments that are unfolded.
One might attempt to perform this procedure for arbitrarily large values of $n$ to reconstruct arbitrarily high moments.
However, $n$ also simultaneously serves as a regularization parameter that restricts the class of generator functions that the algorithm can optimize over.
As $n\to\infty$, this class is the set of all positive analytic functions.
This is a manifestation of the bias--variance trade-off; increasing $n$ to reconstruct higher moments results in a reduction in the precision of the prediction of any individual moment.

\subsection{Adversarial Optimization}
\label{sec:ml_implementation}

To implement Moment Unfolding, we modify the learning setup of a Generative Adversarial Network (GAN)~\cite{Goodfellow:2014upx} to find the optimal values of $\beta_a$ in \Eq{generator}.
As shown schematically in \Fig{schematic}, the weight function $g(z)$ can be viewed as a ``generator'' which is optimized adversarially against a ``discriminator'' that tries to distinguish the reweighted simulation from the experimental data.

In a typical GAN, the generator $g$ surjects a latent space onto a data space, while a discriminator $d$ distinguishes generated examples from real examples.
These two neural networks are then trained simultaneously to optimize the Binary Cross Entropy (BCE) loss functional, where the generator tries to maximize the loss with respect to $g$ while the discriminator tries to minimize the loss with respect to $d$.

For Moment Unfolding, the latent space probability density is the truth-level simulation density, and the generation process is simply reweighting events by $g(z)$, where $g(z)=g_{\rm NN}(z)/\hat{P}$ for neural network $g_{\rm NN}$ and batch-level normalization estimate $\hat{P}$.
Our discriminator is a neutral network $d(x)$ that operates on detector-level distributions.
Instead of BCE, we use the Maximum Likelihood Classifier (MLC)~\cite{DAgnolo:2018cun,DAgnolo:2019vbw,Nachman:2021yvi} loss functional, because it satisfies the analytic closure guarantees proven in \App{analytic}.
That said, we tested BCE for our case studies, finding that it yields similar empirical performance.
The functions $g(z)$ and $d(x)$ are trained simultaneously to optimize a weighted version of MLC loss:
\begin{align}
\label{eq:numericloss}
    L[g,d]=&-\sum_{x\in\mathrm{Data}}\log d(x) \\
    \nonumber&- \sum_{(z,x)\in(\mathrm{Gen},\mathrm{Sim})} g(z) \big(1-d(x)\big),
\end{align}
In the first sum, $x$ are examples obtained from Data, while in the second sum, $(z, x)$ are tuples sampled from Generation and Simulation.
A similar setup (with unrestricted $g$) was used for domain adaptation in \Reff{Erdmann:2020tpv}.

For the empirical studies in \Sec{case}, each neural network is implemented using \textsc{Keras}~\cite{keras} with the \textsc{Tensorflow2} backend~\cite{tensorflow} and optimized with \textsc{Adam}~\cite{adam}.
The discriminator function $d$ has three hidden layers, using 50 nodes per layer.
Rectified Linear Unit (ReLU) activation functions are used for the intermediate layers and a sigmoid function is used for the last layer.

\section{Case Studies}
\label{sec:case}

To demonstrate the features of Moment Unfolding, we perform numerical cases studies, both in a Gaussian example and in the realistic setting of jet measurements at the Large Hadron Collider (LHC).

\subsection{Gaussian Example}

\begin{figure*}
    \centering
    \subfloat[]{\includegraphics[height=0.32\textwidth]{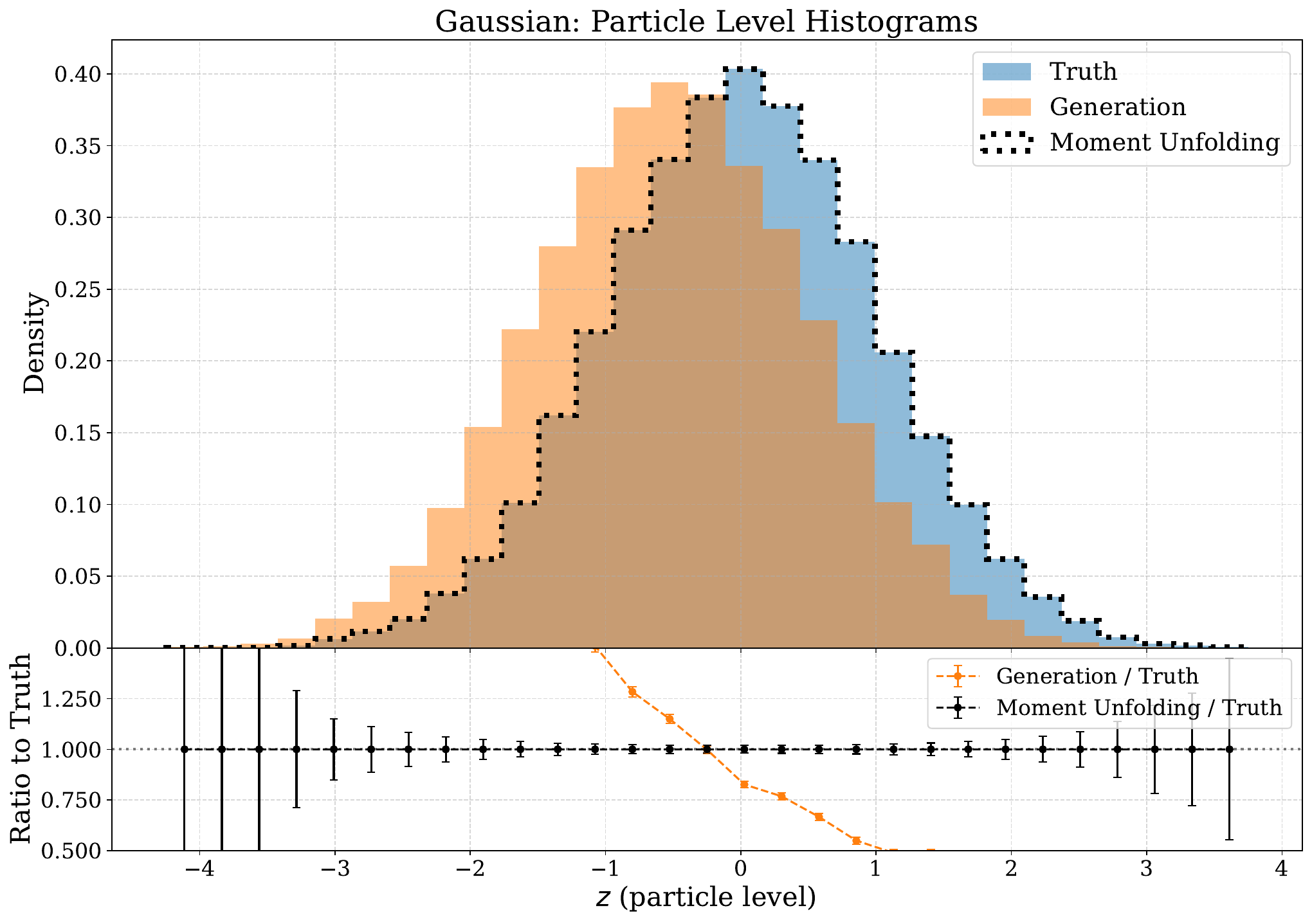}
    \label{fig:gaussexample}}
    $\qquad$
    \subfloat[]{\includegraphics[height=0.32\textwidth]{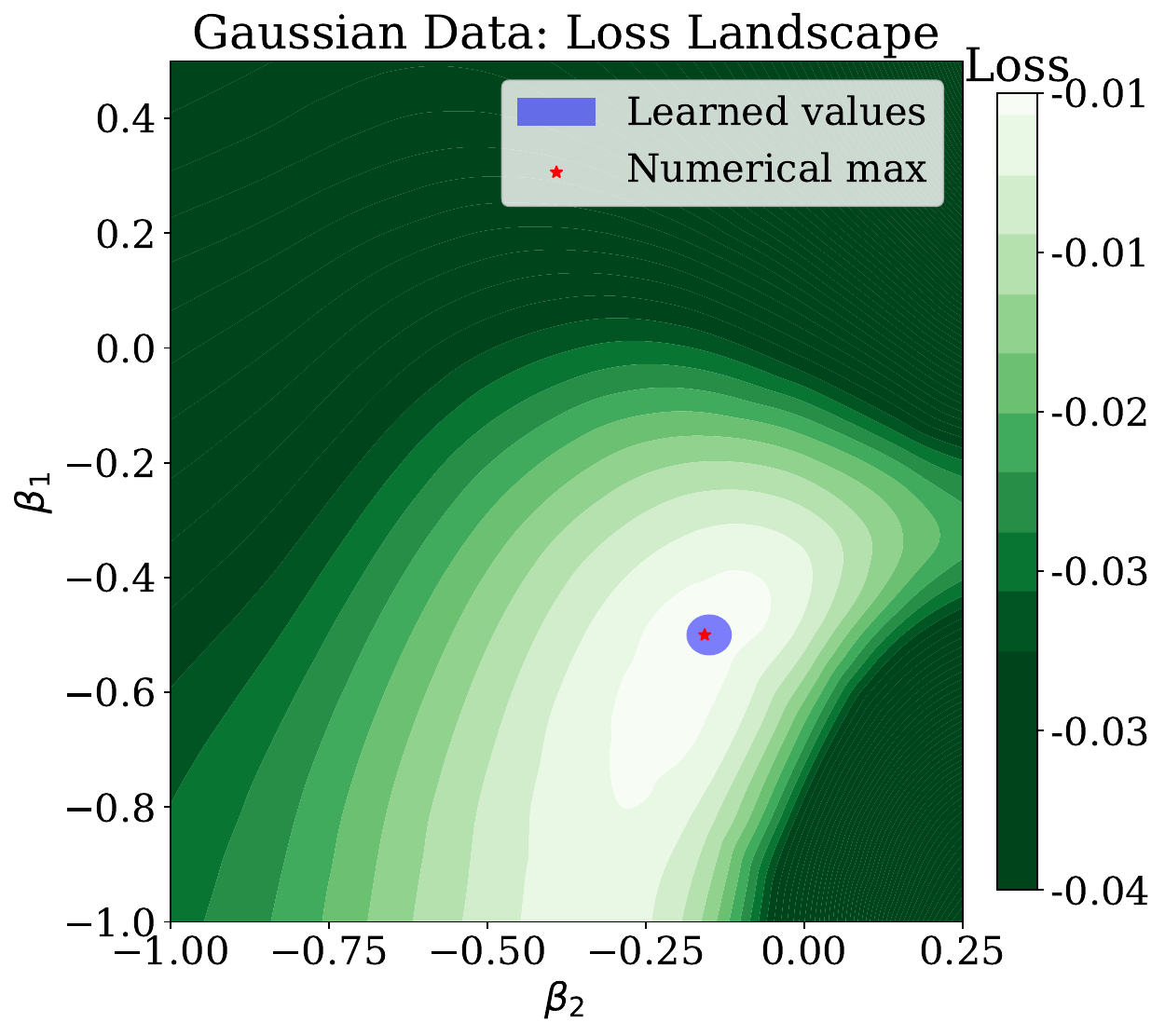}
        \label{fig:gaussloss}}
    \caption{
    (i) Distributions from the Gaussian example of particle-level truth, generation, and reweighted generation (i.e.~Moment Unfolding).
    The agreement between the truth and reweighted samples demonstrates the qualitative performance of Moment Unfolding.
    (ii) The weighted MLC loss from \Eq{numericloss} for fixed $g$ but optimized $d$, found by scanning over $\beta_1$ and $\beta_2$.
    The correct value is indicated by a red star.
    Indicated in shaded blue is the $1\sigma$ bootstrapped interval for Moment Unfolding's prediction of $\beta_a$.
        }
        \label{fig:gauss}
\end{figure*}

We begin by unfolding Gaussian data with Gaussian distortions.
Let $\N(\mu, \sigma^2)$ be a normal distribution with mean $\mu$ and variance $\sigma^2$.
The particle-level truth distribution is drawn from $\N(0, 1)$ while the particle-level generation is drawn from $\N(-0.5, 1)$.
Detector effects are represented by additive noise that is also Gaussian, with distribution $\N(0, 5)$.
Since the Gaussian probability density is uniquely specified by its first two moments, Moment Unfolding with weighting function
\begin{equation}
\label{eq:weight_for_gaussian}
g(z) = \frac1Pe^{-\beta_1 z - \beta_2z^2}
\end{equation}
can in principle result in a perfect unfolding of the entire distribution.
As discussed in \App{analytic}, no unfolding method can be successful in all cases, and Moment Unfolding in particular cannot recover the true distribution when there are large off-diagonal elements in the detector response, which is indeed the case here.

Nevertheless, as shown in \Fig{gaussexample}, the numerical results of applying Moment Unfolding to this Gaussian example are quite promising.
Here, we show histograms at particle level, comparing the truth dataset (blue shaded), generation dataset (orange shaded), and the result of weighting the generation dataset by \Eq{weight_for_gaussian} (black dotted line).
Visually, the close overlap between the truth histogram and the weighted generation histogram shows that the Moment Unfolding procedure was successful.

Since this is a simple one-dimensional problem, we can study the success of this procedure more precisely.
Specifically, we can check whether the maximum of the discriminator loss function is indeed at values of $\beta_a$ that correspond to the moments of the truth distribution.
In practice, we do not have access to the full loss landscape, but for this one-dimensional problem, we can scan over a discrete set of generator parameters for illustration.
Then, to obtain the value of the loss, we can train the discriminator for fixed values of $\beta_a$.

In \Fig{gaussloss}, we plot the discriminator-optimized loss as a function of $\beta_1$ and $\beta_2$.
The red star represents the loss maximum.
The solid blue region represents the $1\sigma$ confidence interval for the values of $\beta_a$ learned by the Moment Unfolding algorithm, estimated from a bootstrapping procedure.
Since the red star coincides with the corresponding solid blue ellipse, the success of the procedure is verified.

\subsection{Jet Substructure}

\setlength{\tabcolsep}{6pt}
\renewcommand{\arraystretch}{1.5}
\begin{table*}[p]
\caption{Moments of jet observables at particle level. The uncertainties in the Truth and Generation columns are computed by bootstrapping the datasets. The uncertainties in the Moment Unfolding column are computed by adding the uncertainty in Generation in quadrature to the empirical uncertainty obtained by computing the $1\sigma$ confidence interval for the moment predicted by Moment Unfolding on the same dataset multiple times.}
\begin{tabular}{ p{.1\textwidth} p{.2\textwidth} p{.2\textwidth} p{.2\textwidth} }
\hline
Observable& Truth &Generation & Moment Unfolding \\
\hline
\hline
$\ev{M}$ & $(2.182\pm 0.030)\times 10^1$ & $(2.064\pm 0.043)\times 10^1$ & $(2.173 \pm 0.047)\times 10^1$\\
$\ev{M^2}$ & $(6.049\pm 0.222)\times 10^2$ & $(5.360\pm0.350)\times 10^2$ & $(6.115\pm0.364)\times 10^2$  \\
\hline
$\ev Q$ & $(1.006\pm 0.037)\times 10^{-2}$ & $(1.582 \pm 0.038)\times 10^{-2}$ & $(1.090\pm 0.040)\times10^{-2}$ \\
$\ev{Q^2}$    & $(1.216\pm 0.082)\times10^{-2}$ & $(1.508\pm 0.074)\times 10^{-2}$ & $(1.207\pm 0.074)\times 10^{-2}$ \\
\hline
$\ev W$ & $(1.498\pm 0.025) \times 10\inv$ &$(1.231\pm 0.029)\times 10\inv$ & $(1.499\pm0.029)\times10\inv$ \\
$\ev{W^2}$ & $(3.370\pm 0.113) \times 10^{-2}$   & $(2.421\pm 0.128)\times 10^{-2}$ & $(3.374\pm 0.128)\times10^{-2}$ \\
\hline
$\ev{Z_g}$ & $(2.334\pm 0.029)\times10^{-1}$& $(2.457\pm 0.030)\times 10\inv$ & $(2.353\pm 0.059)\times10\inv$\\
$\ev{Z_g^2}$ & $(6.789\pm 0.166)\times10^{-2}$ & $(7.425\pm 0.165)\times10^{-2}$ & $(6.767\pm 0.330)\times10^{-2}$
\\
\hline
\hline
\end{tabular}
\label{tab:jetmoments}
\end{table*}

\begin{figure*}[p]
    \subfloat[]{\includegraphics[height=0.28\textwidth]{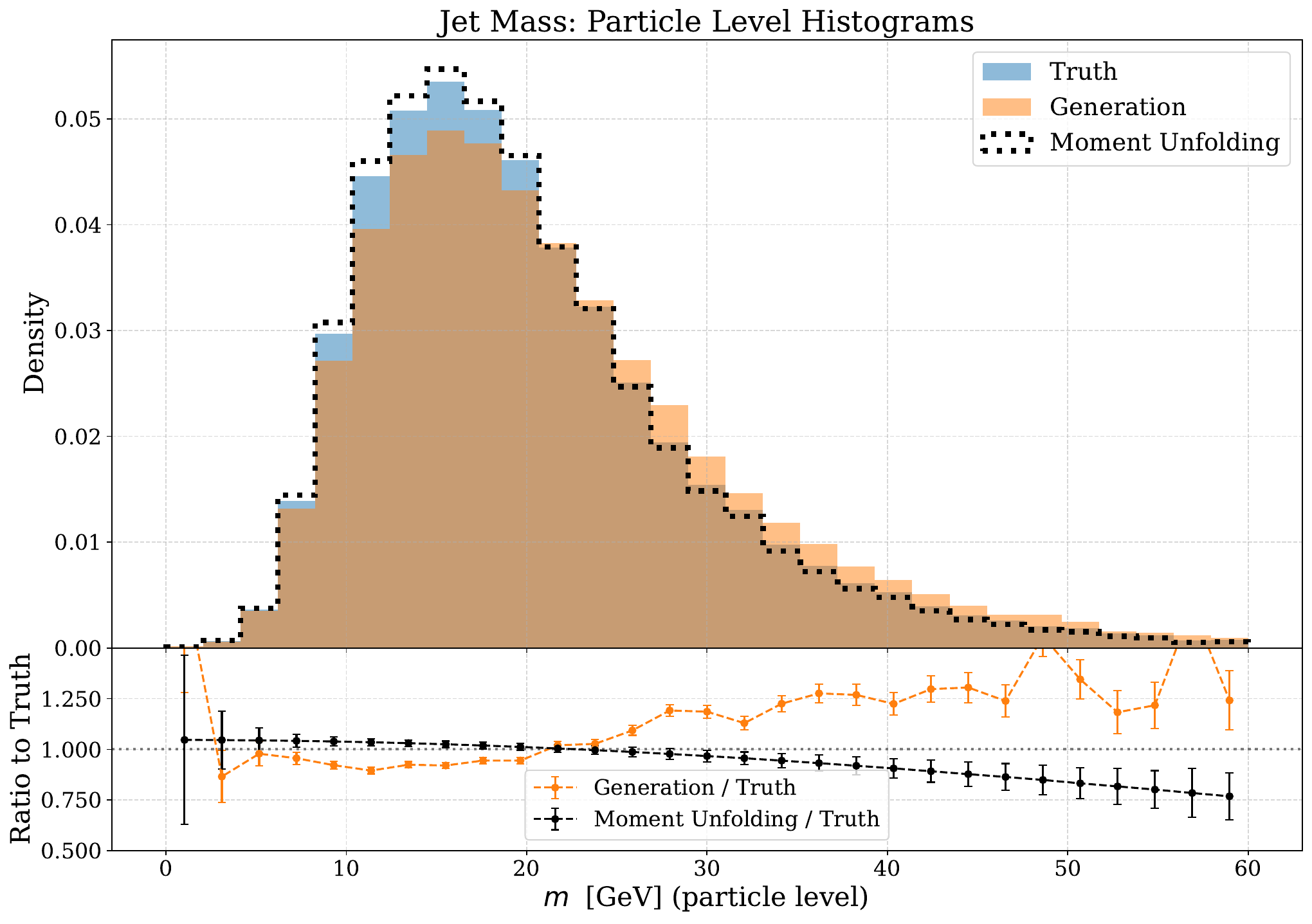}}
    $\qquad$
    \subfloat[]{\includegraphics[height=0.28\textwidth]{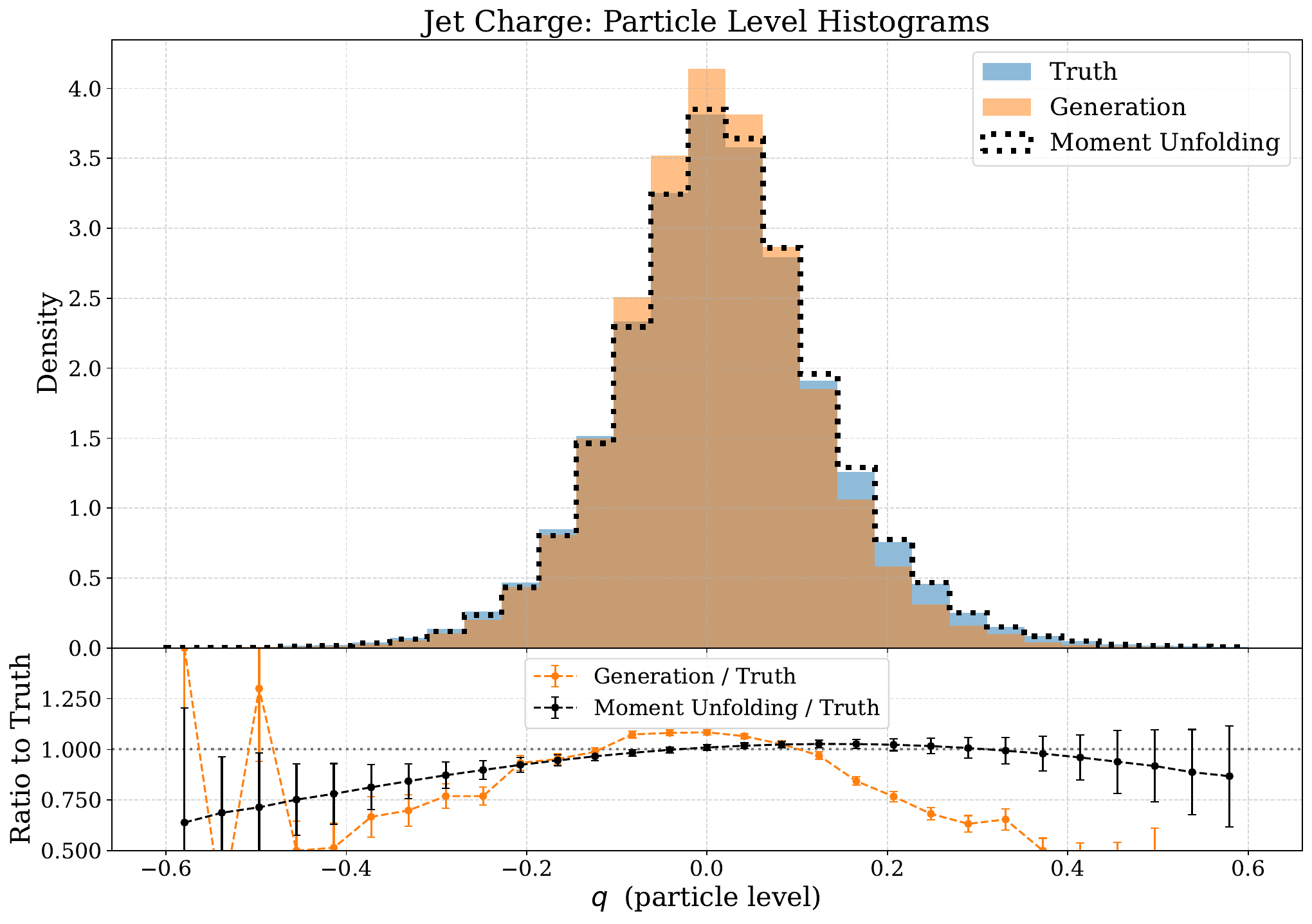}}\\

    \subfloat[]{\includegraphics[height=0.28\textwidth]{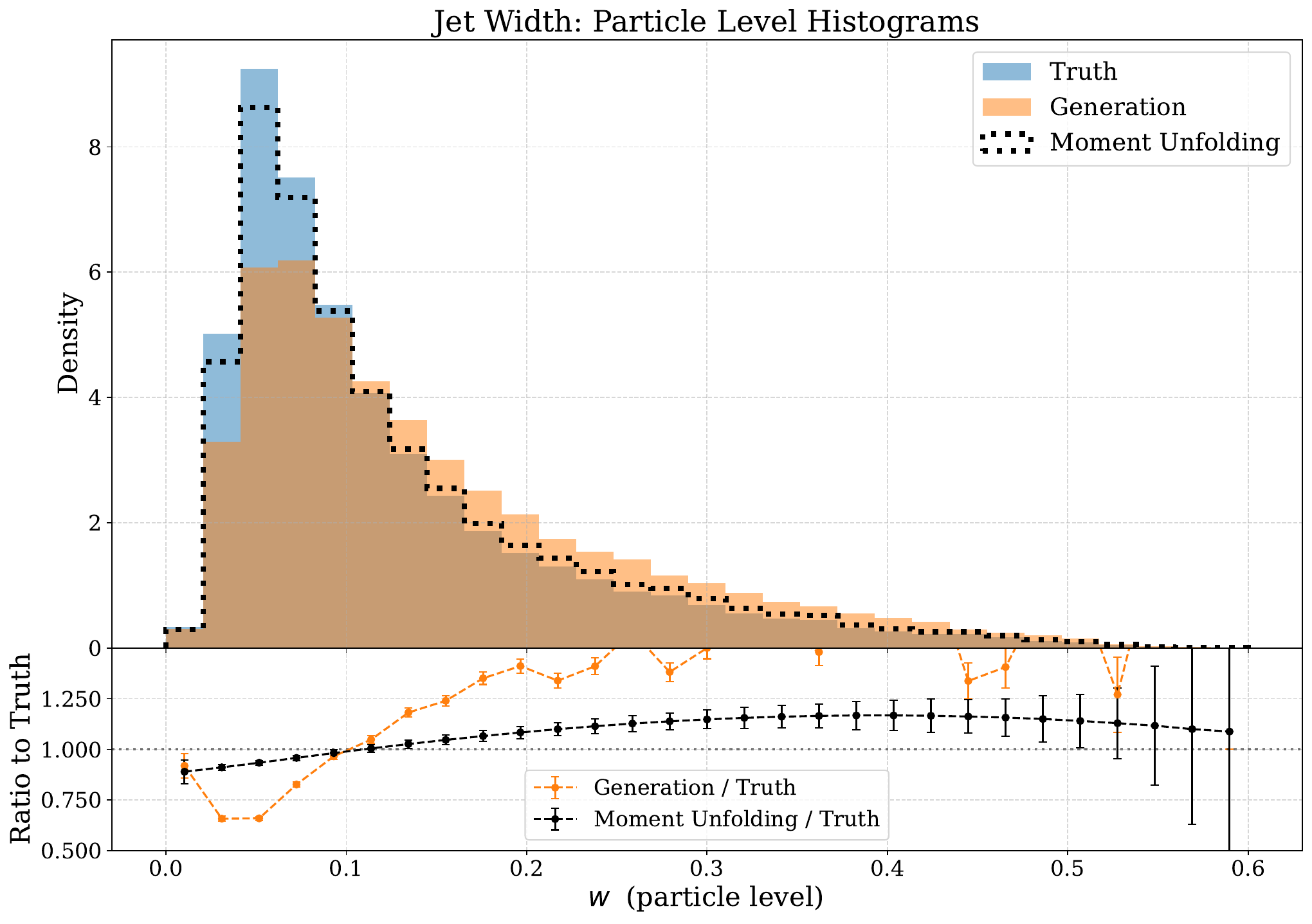}}
    $\qquad$
    \subfloat[]{\includegraphics[height=0.28\textwidth]{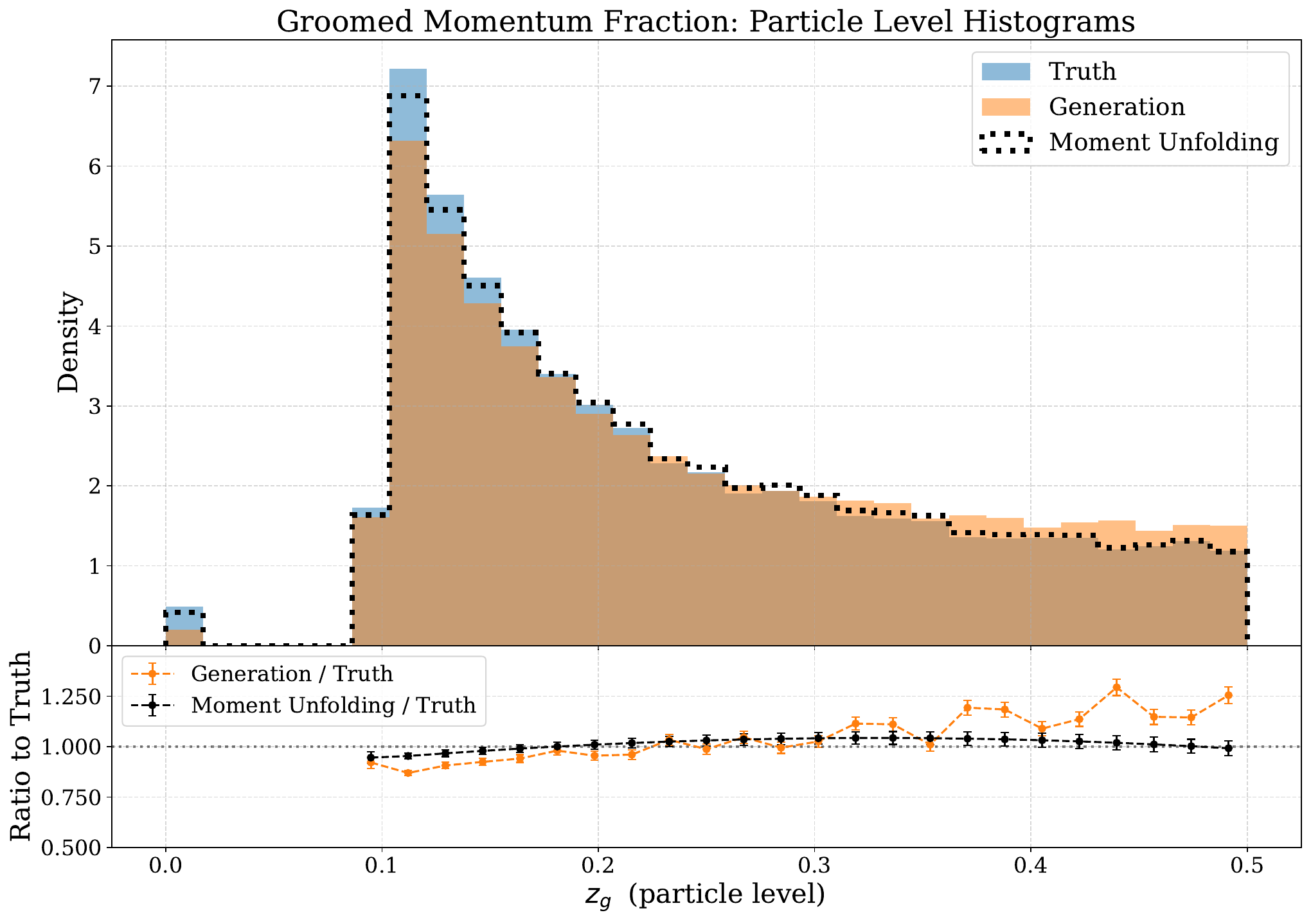}}
    \caption{
    Distributions of (i) jet mass, (ii) jet charge, (iii) jet width, and (iv) groomed momentum fraction at particle-level, comparing truth, generation, and the results from Moment Unfolding.
    }
    \label{fig:jetexample_dists}
\end{figure*}

\begin{figure*}[t]
    \subfloat[]{ \includegraphics[height=0.32\textwidth]{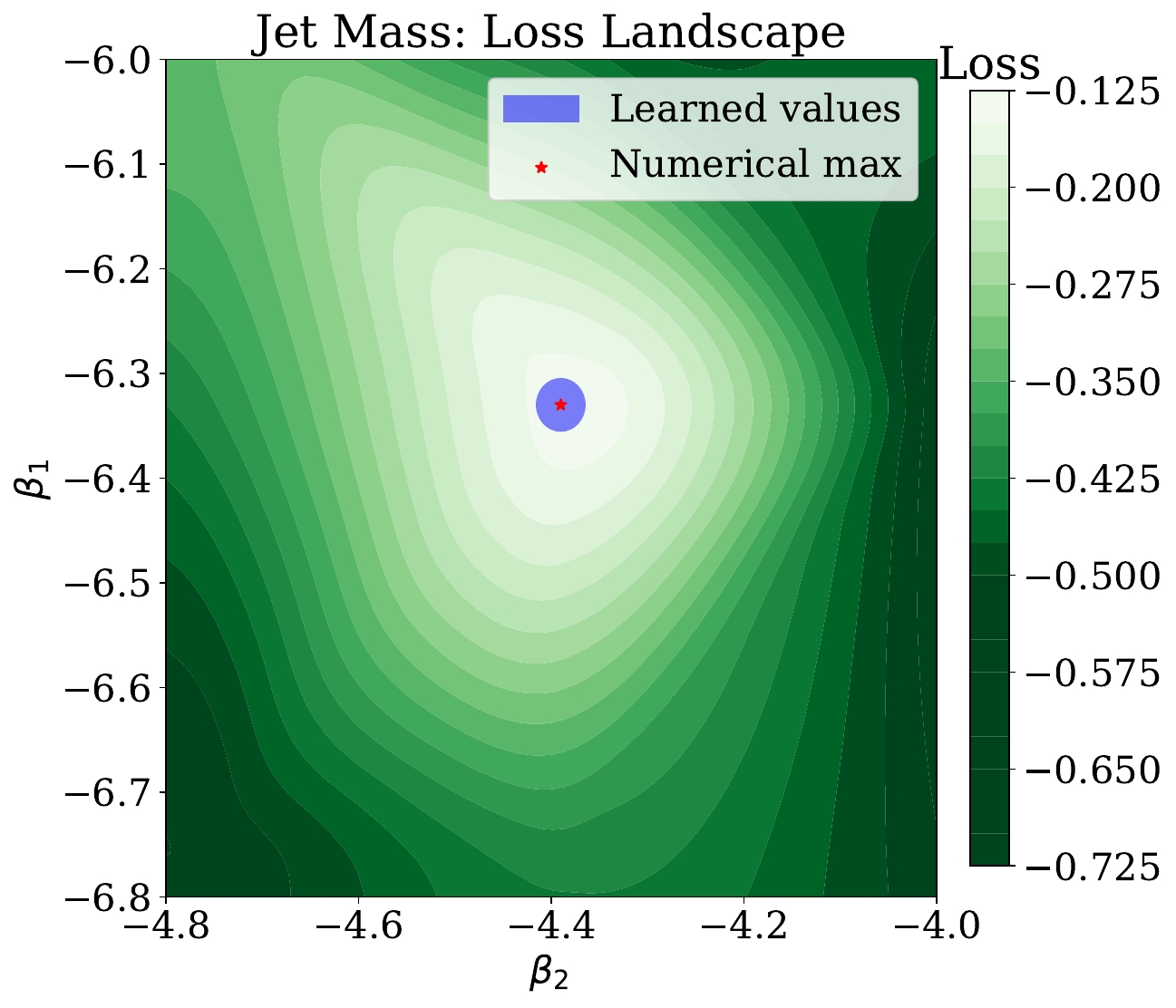}\label{fig:mjetloss}}
    $\qquad$
    \subfloat[]{ \includegraphics[height=0.32\textwidth]{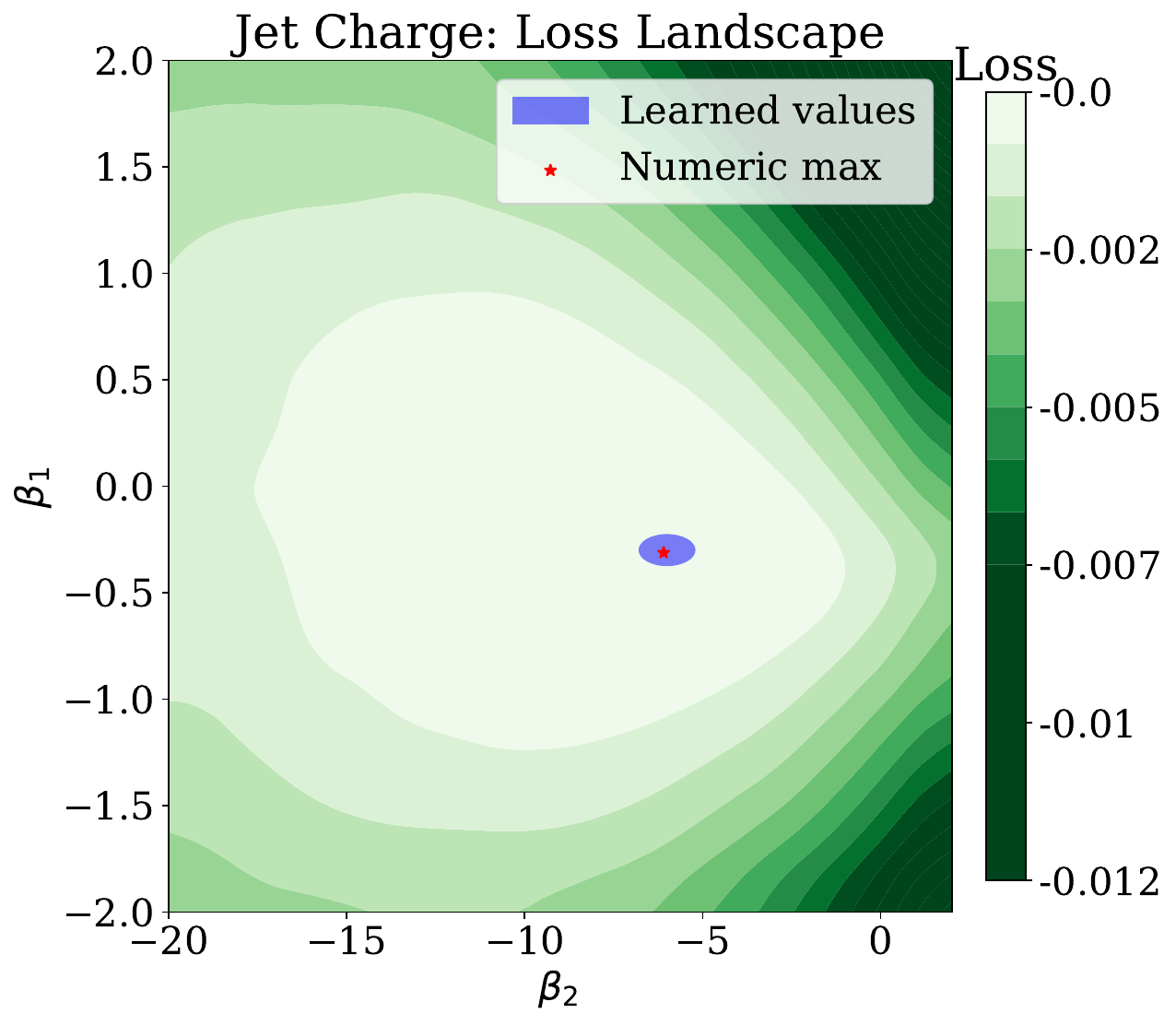}\label{fig:qjetloss}}\\
    
    \subfloat[]{ \includegraphics[height=0.32\textwidth]{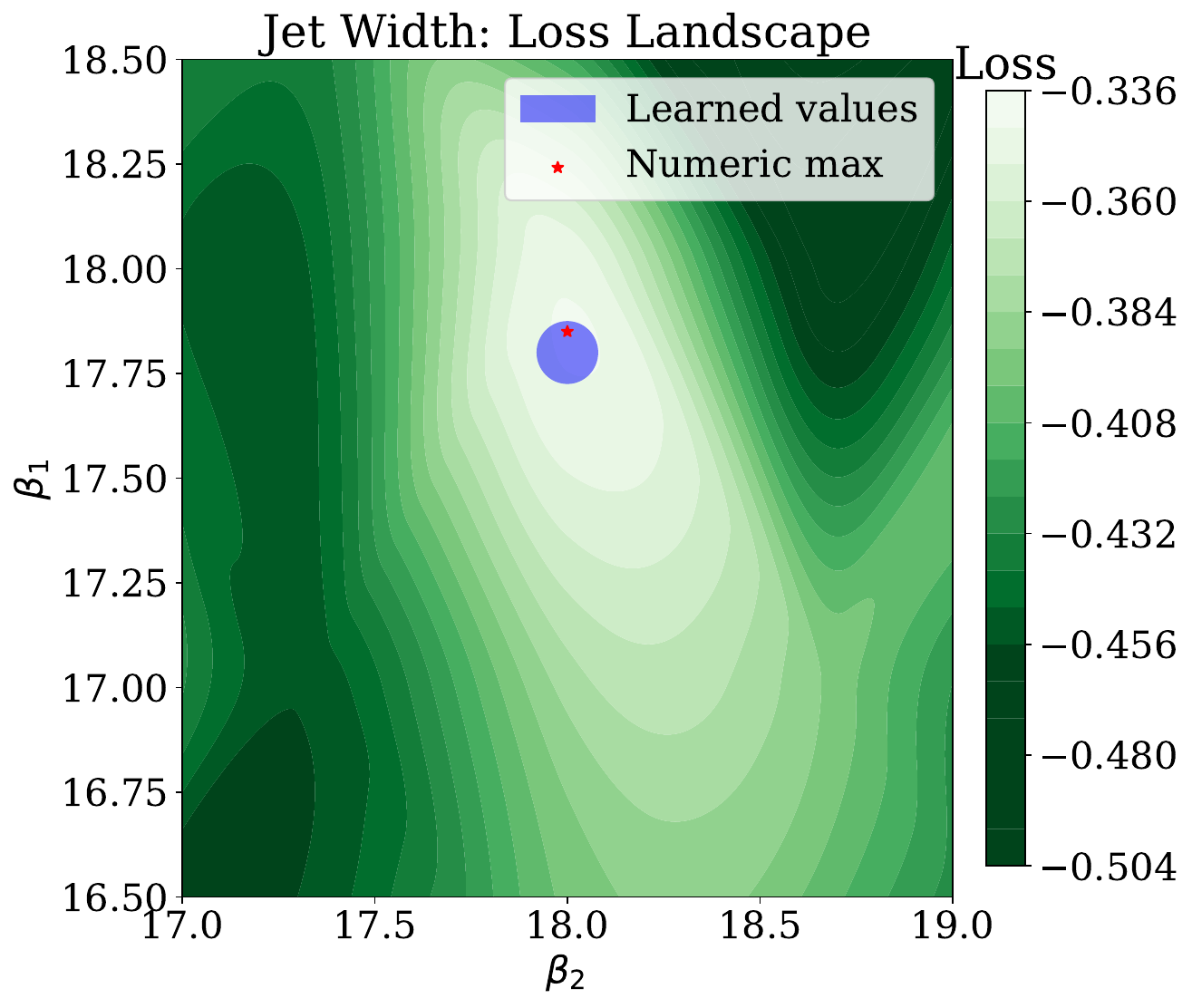}\label{fig:wjetloss}}
    $\qquad$
    \subfloat[]{ \includegraphics[height=0.32\textwidth]{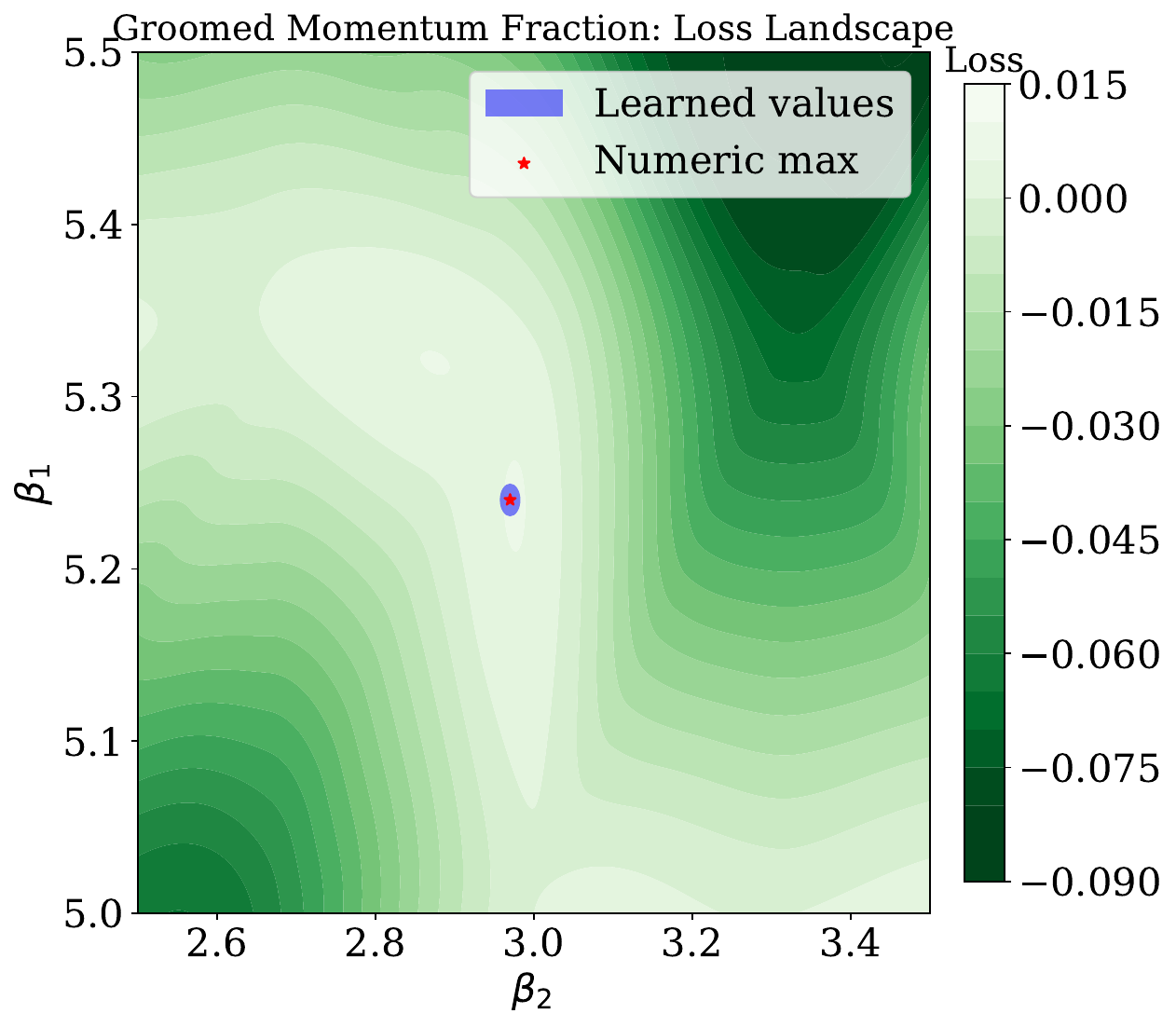}\label{fig:zgjetloss}}
    \caption{
    The discriminator-optimized MLC loss as a function of $\beta_1$ and $\beta_2$, for (i) jet mass, (ii) jet charge, (iii) jet width, and (iv) groomed momentum fraction.
    The correct $\beta_a$ values are indicated by red dot, while the $1\sigma$ intervals for Moment Unfolding's predictions of $\beta_a$ are shown as a blue circle.
    }
    \label{fig:jetexample_loss}
\end{figure*}

As a more realistic case study, we study hadronic jets from the LHC. 
Jets are collimated sprays of particles that arise from the fragmentation of high-energy quarks and gluons.
Measuring the substructure of jets is an active area of research, both to understand QCD dynamics and to search for physics beyond the Standard Model~\cite{Larkoski:2017jix,Kogler:2018hem}.
For this study, we consider four jet substructure observables: jet mass $m$, jet charge $q$ with $\kappa = 1/2$~\cite{Krohn:2012fg}, jet width $w$~\cite{Larkoski:2017jix}, and momentum fraction $z_g$~\cite{Larkoski:2015lea} after Soft Drop grooming~\cite{Dasgupta:2013ihk,Larkoski:2014wba} with $z_\mathrm{cut} = 0.1$ and $\beta = 0$:
\begin{align}
m&=\sqrt{\sum_k E_k^2-\sum_k p_k^2}\,,\\
q&=\frac{1}{\sum_k \sqrt{p_{T,k}}}\sum_k q_k\,\sqrt{p_{T,k}}\,,\\
w&=\frac{1}{p_{T,\text{jet}}}\sum_k p_{T,k} \, \Delta R_k\,,\\
z_g &= \frac{p_{T,\mathrm{subleading}}}{p_{T,\mathrm{leading}} + p_{T,\mathrm{subleading}}}.
\end{align}
Here, the sums run over the constituents of a jet, and $E_k$, $p_k$, $p_{T,k}$, $\Delta R_k$, and $q_k$ are the constituent energy, three-momentum, transverse momentum, angular distance from the jet axis, and electric charge for particle $k$, respectively, and $p_{T,\mathrm{(sub)leading}}$ is the transverse momentum of the (sub)leading prong~\cite{ALargeIonColliderExperiment:2021mqf} returned by the Soft Drop algorithm.
The jet mass and jet width are examples of infrared-and-collinear-safe observables that are expected to be less sensitive to detector effects, while the groomed momentum fraction (as $\beta \to 0$) is Sudakov safe~\cite{Larkoski:2015lea}.
Jet charge is infrared but not collinear safe and therefore expected to be more susceptible to certain kinds of detector distortions.
The first moment of the jet charge distribution as a function of jet $p_T$ has been calculated in \Reffs{Krohn:2012fg,Waalewijn:2012sv,Li:2019dre,Kang:2021ryr,Kang:2020fka} and measured by ATLAS~\cite{ATLAS:2015rlw} and CMS~\cite{CMS:2017yer,CMS:2020plq}.

The simulated samples used for this study are the same as in \Reffs{Andreassen:2019cjw,andreassen_2019_3548091} and briefly summarized here.
Proton-proton collisions are simulated at $\sqrt{s}=14$ TeV with the default tune of Herwig~7.1.5~\cite{Bahr:2008pv,Bellm:2015jjp,Bellm:2017bvx} and Tune 26~\cite{ATL-PHYS-PUB-2014-021} of Pythia~8.243~\cite{Sjostrand:2007gs,Sjostrand:2006za,Sjostrand:2014zea}.
As a proxy for detector effects and a full detector simulation, we use the Delphes~3.4.2~\cite{deFavereau:2013fsa} fast simulation of the CMS detector, which uses particle flow reconstruction~\cite{CMS:2017yfk}.
Jets with radius parameter $R=0.4$ are clustered using either all particle flow objects (detector-level) or stable non-neutrino truth particles (particle-level) with the anti-$k_T$ algorithm~\cite{Cacciari:2008gp} implemented in FastJet~3.3.2~\cite{Cacciari:2011ma,Cacciari:2005hq}.
To reduce acceptance effects, the leading jets are studied in events with a $Z$ boson with transverse momentum $p_T^Z>200$~GeV.
For this study, we treat Pythia+Delphes as ``truth/data" and Herwig+Delphes as ``generation/simulation".

For each of the four observables above, we unfold the first two moments from detector-distorted jet substructure data.
We emphasize that this unfolding is done separately for each observable, leaving joint unfolding to future work.
The moment results are presented in \Tab{jetmoments}, where we see that Moment Unfolding has good performance recovering the expected truth moments within statistical uncertainties.
The uncertainties in the ``Truth'' and ``Generation'' columns are computed by bootstrapping the respective datasets, computing the relevant moment for each bootstrapped dataset, and then computing the $1\sigma$ interval for this moment. The uncertainties in the ``Moment Unfolding'' column are computed by adding the uncertainty in ``Generation'' in quadrature to the intrinsic uncertainty of the empirical procedure estimated by learning the same moment on the same dataset multiple times and computing the $1\sigma$ confidence interval for the predicted moment.

As an alternative visualization, we show the inferred particle-level distributions in \Fig{jetexample_dists}.
The jet mass follows a unimodal distribution peaked at about 20 GeV, with a relatively long tail many standard deviations to the right.
These features are observed both in  truth and generation with a sharper peak in truth.
The jet charge has an approximately Gaussian distribution, and is close to symmetric with a small positive skew because approximately equal number of positively- and negatively-charged particles are produced, with a small excess of positively charged particles produced because this is a proton-proton process.
Radiation within jets is enhanced at low values of $\Delta R$ which leads to the uni-modal distribution of the jet width that falls off rapidly after about 0.1; these features are present in both the truth and generation, albeit with a longer tail in truth.
The groomed momentum fraction offers an opportunity to study the performance of Moment Unfolding for data that is not well approximated as a Gaussian distribution and has a sharp cutoff feature.
For all four observables, even though the first and second moments of the reweighted generation match the truth well, the full distributions are not statistically identical.
This is because higher moments are relevant and are not the same between truth and generation.

In \Fig{jetexample_loss}, we perform
a loss function analysis where we scan over values of $\beta_a$, learn the optimal discriminator for the fixed generator, and then compare the loss function maximum to the learned values of $\beta_a$.
The red star represents the maximum of the MLC loss.
The solid blue region represents the $1\sigma$ confidence interval for the values of $\beta_a$ learned by the Moment Unfolding algorithm, estimated from a bootstrapping procedure.
Since the red star coincides with the corresponding solid blue ellipse, the success of the procedure is verified.

\begin{figure*}[t]
    \centering
    \subfloat[]{\includegraphics[height=0.28\textwidth]{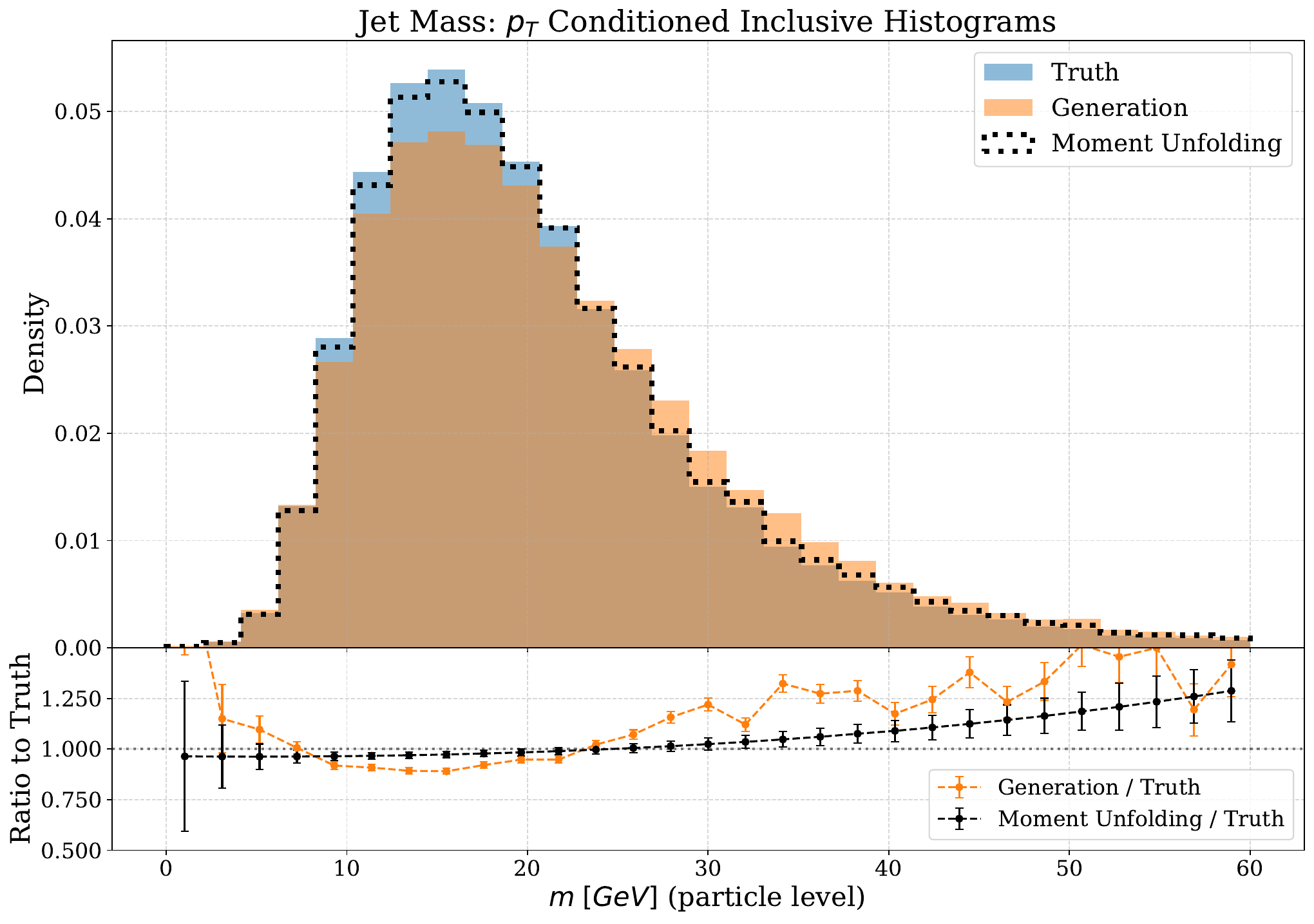}}
    $\qquad$
    \subfloat[]{\includegraphics[height=0.28\textwidth]{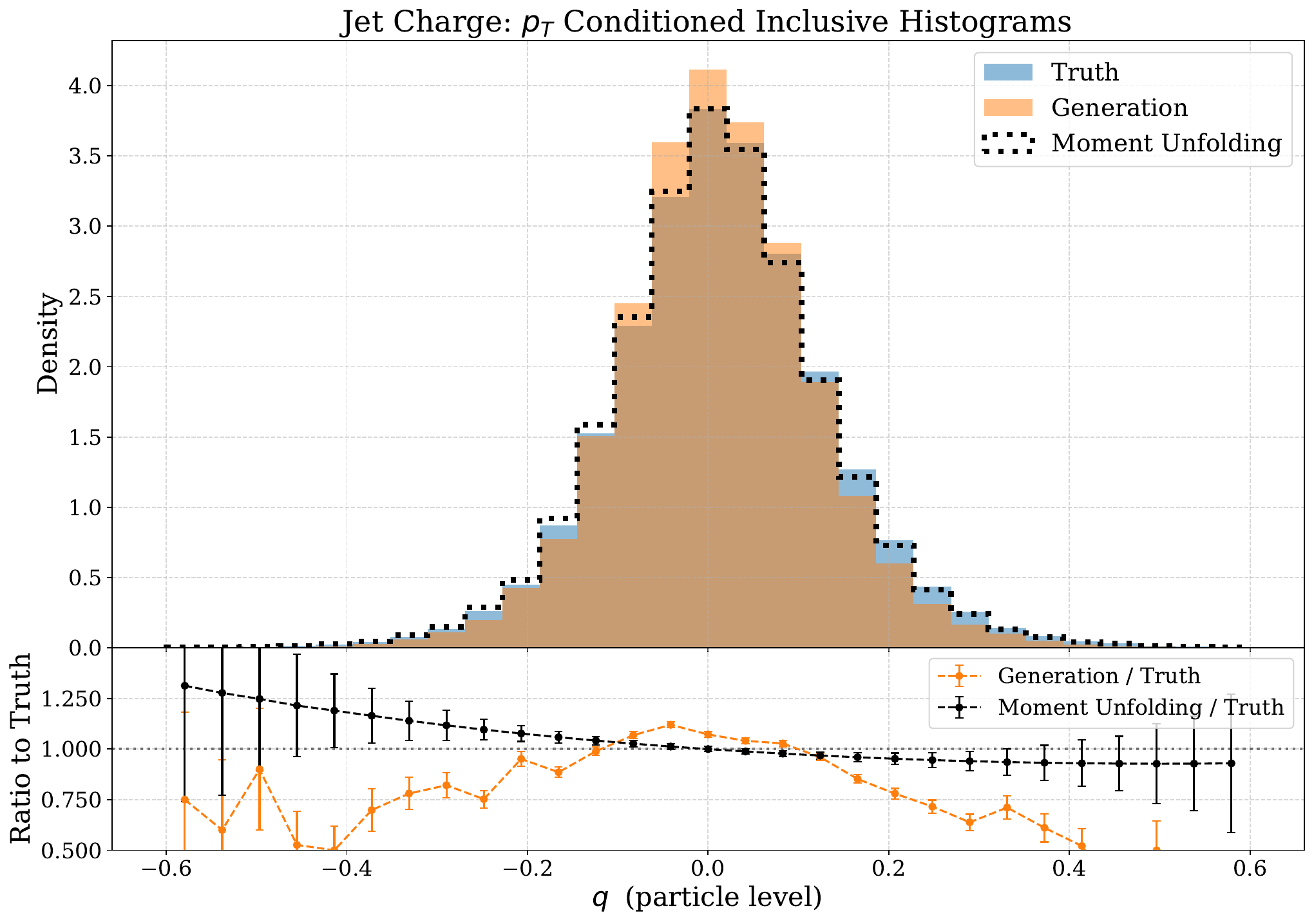}}\\
    \subfloat[]{\includegraphics[height=0.28\textwidth]{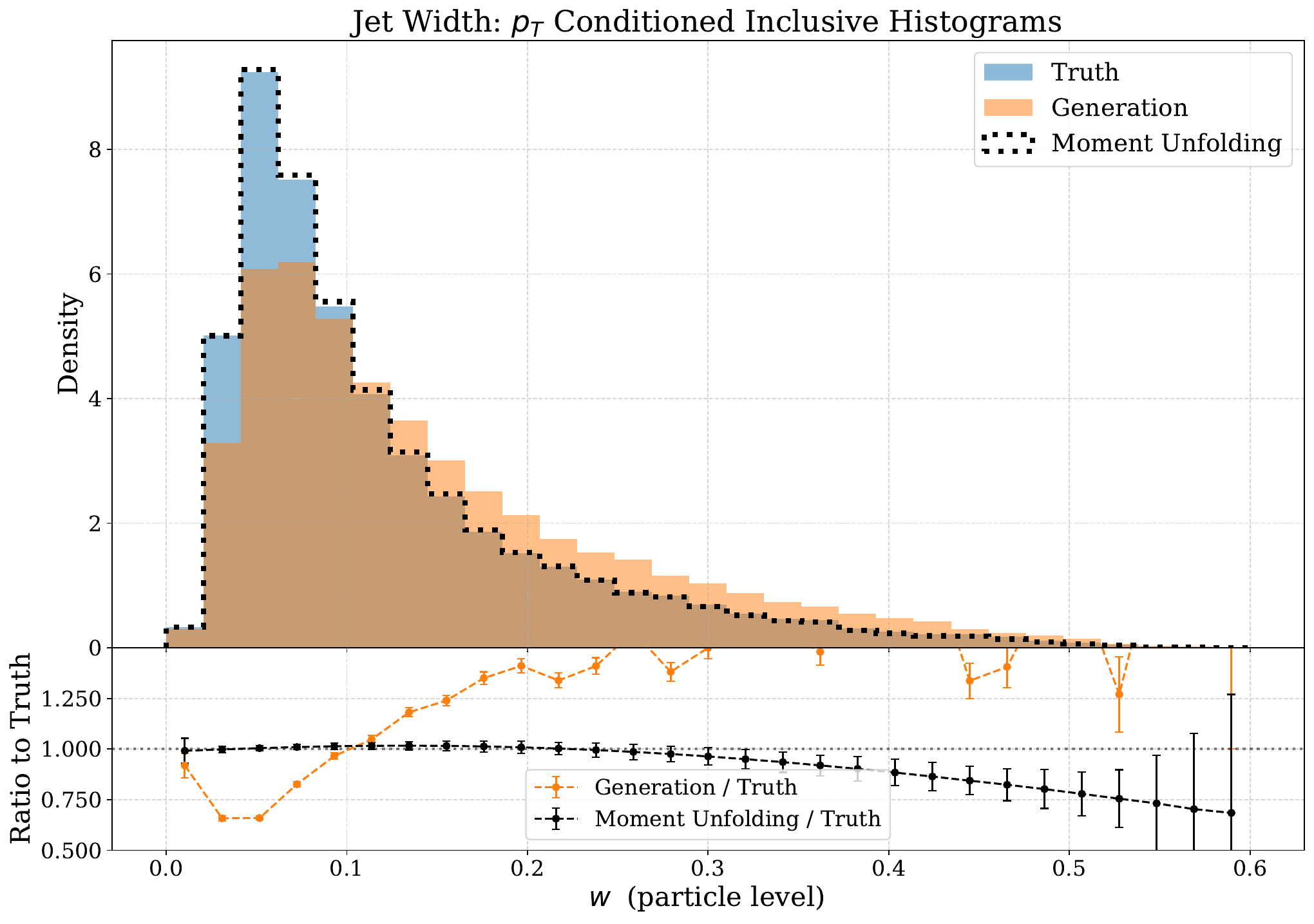}}
    $\qquad$
    \subfloat[]{\includegraphics[height=0.28\textwidth]{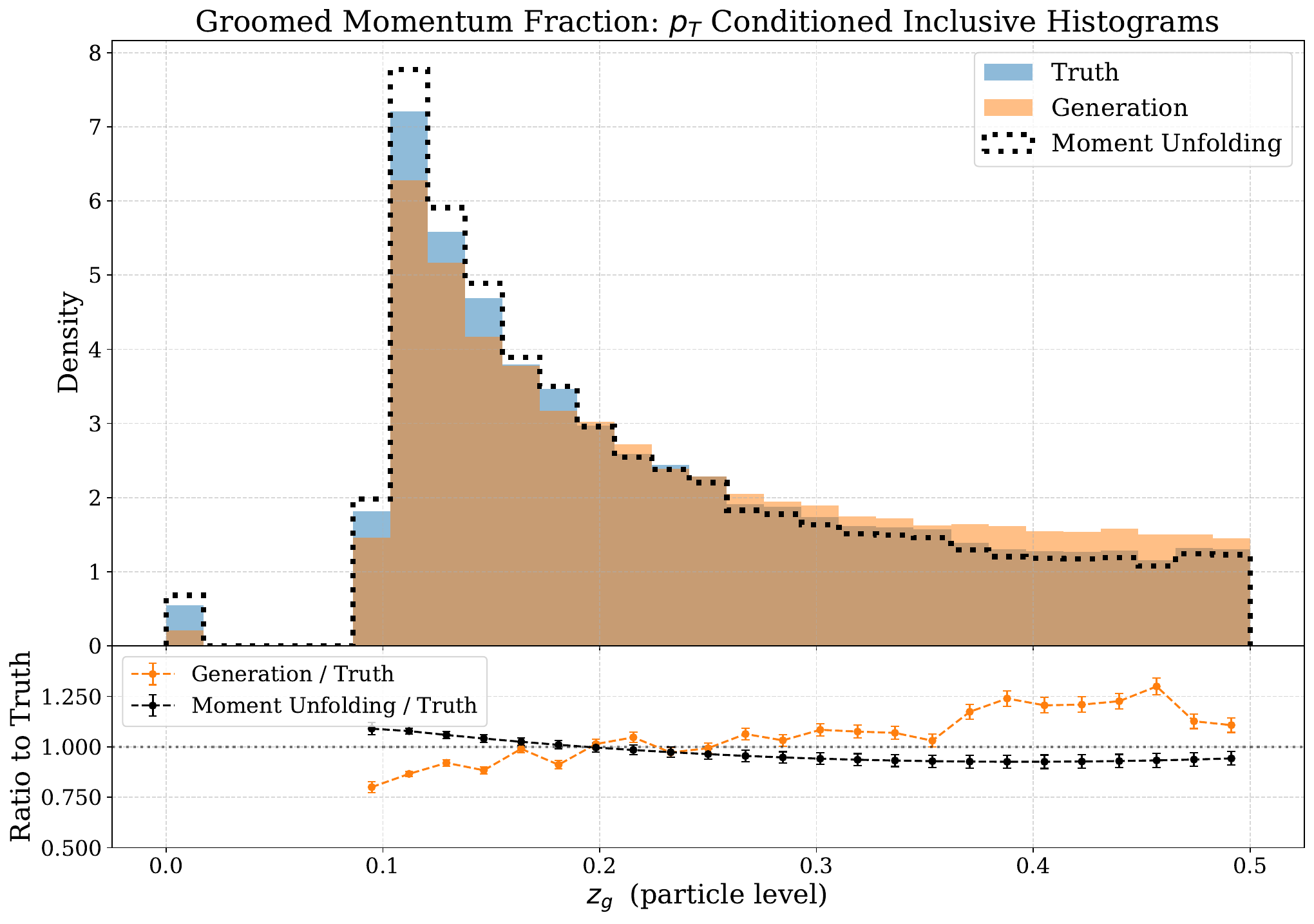}}
    \caption{
    Inclusive distributions of (i) jet mass, (ii) jet charge, (iii) jet width, and (iv) groomed momentum fraction unfolded conditionally on the traverse momentum of the jet.
    Compared to the unconditional unfolding in \Fig{jetexample_dists}, the agreement between truth and Moment Unfolding is typically better.}
    \label{fig:pjetexample}
\end{figure*}

\subsection{Momentum Dependence}
\label{sec:momentum_dependence}

Typically, we are interested in more than just inclusive moments.
For jet substructure observables, it is interesting to study the moments as a function of jet $p_T$.
We can slightly modify our generator $g$ to accommodate this case by adding momentum dependence to the coefficients in \Eq{generator}:
\begin{equation}
\label{eq:momentum_dependent_weight}
    g(z ; p_T) = \frac1P \exp \Bigg[-\sum_{a = 1}^n\beta_a(p_T)\,z^a \Bigg]\,,
\end{equation}
where $\beta_a(p_T)$ is an arbitrary function of $p_T$.
While we could parametrize $\beta_a(p_T)$ as a neural network, we found that this resulted in unstable performance.
Since the $p_T$-dependence is often weak and since the starting simulations are already quite accurate, we regularize the $\beta_a$ by parametrizing them as low-order polynomials.
Empirically, we observe that the ratio of spectra between Pythia and Herwig is approximately linear, so we restrict the $\beta_a$ to be first-order polynomials in $p_T$:
\begin{equation}
    \beta_a (p_T) = \beta_a^{(0)} + \beta_a^{(1)}\, p_T\,.
\end{equation}

In \Fig{pjetexample}, we show the results of carrying out this procedure for the jet mass, jet charge, jet width, and groomed momentum fraction on the inclusive distribution.
These can be contrasted with the plots in \Fig{jetexample_dists} which did not have $p_T$ dependence in the unfolding.
With a $p_T$-dependent weight function, there is stronger similarity between the truth and Moment Unfolding.
Having observed the improved inclusive behavior, we now study jet observable moments differentially in $p_T$.

\begin{figure*}
    \centering
    \subfloat[]{\includegraphics[height=0.25\textwidth]{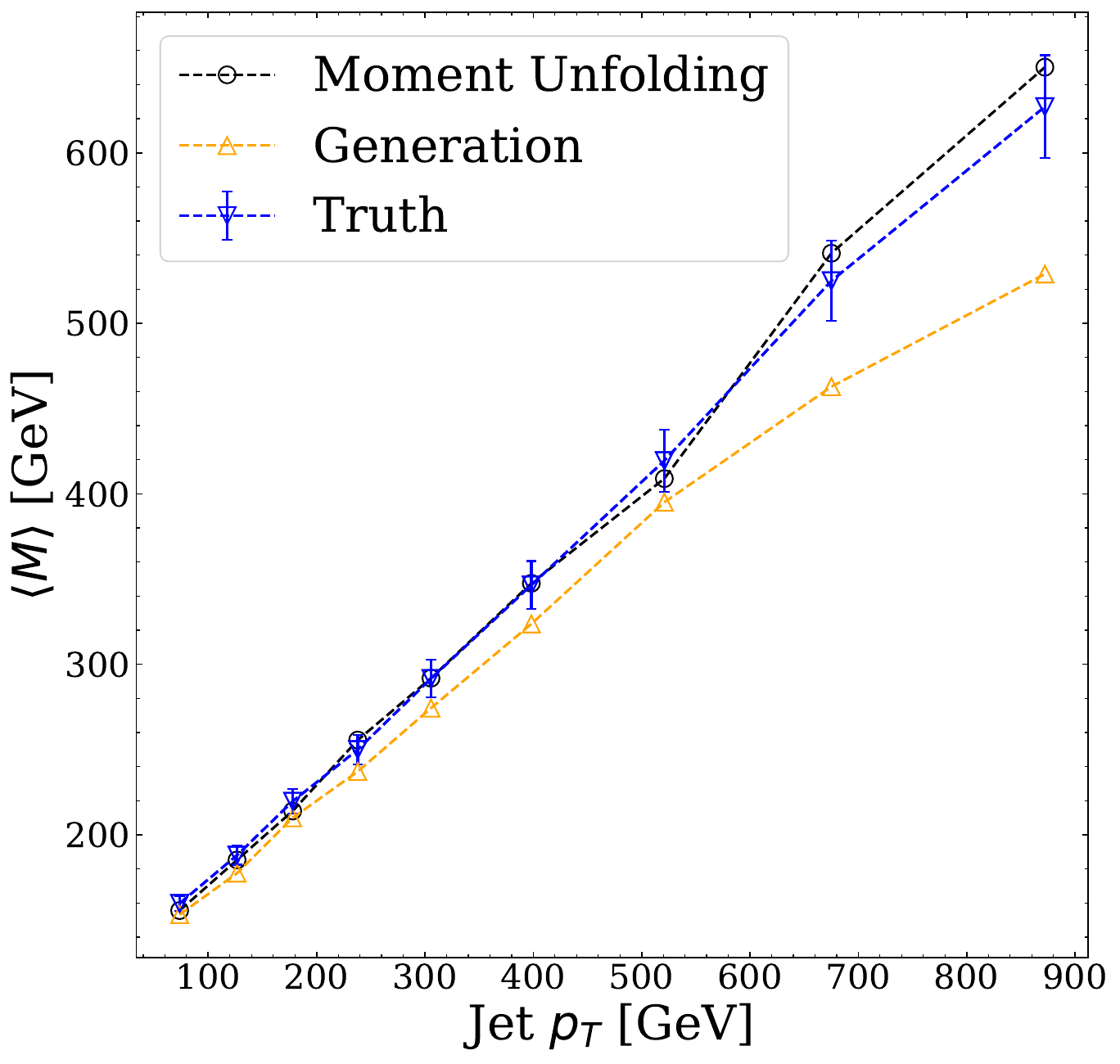}
    $\qquad\qquad$
    \label{fig:pmjetmean}}
\subfloat[]{\includegraphics[height=0.25\textwidth]{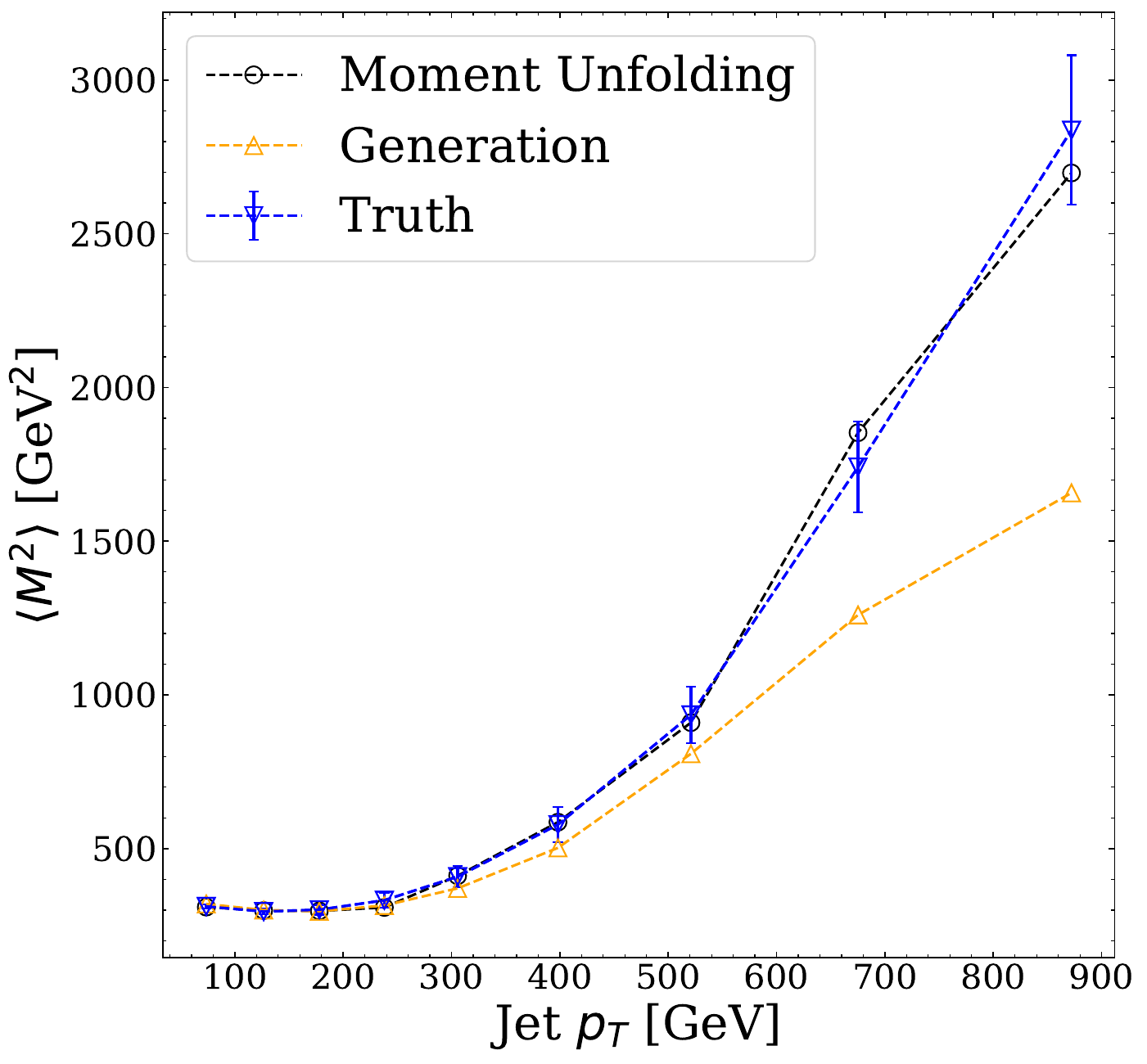}
    \label{fig:pmjetvar}}

        \subfloat[]{\includegraphics[height=0.25\textwidth]{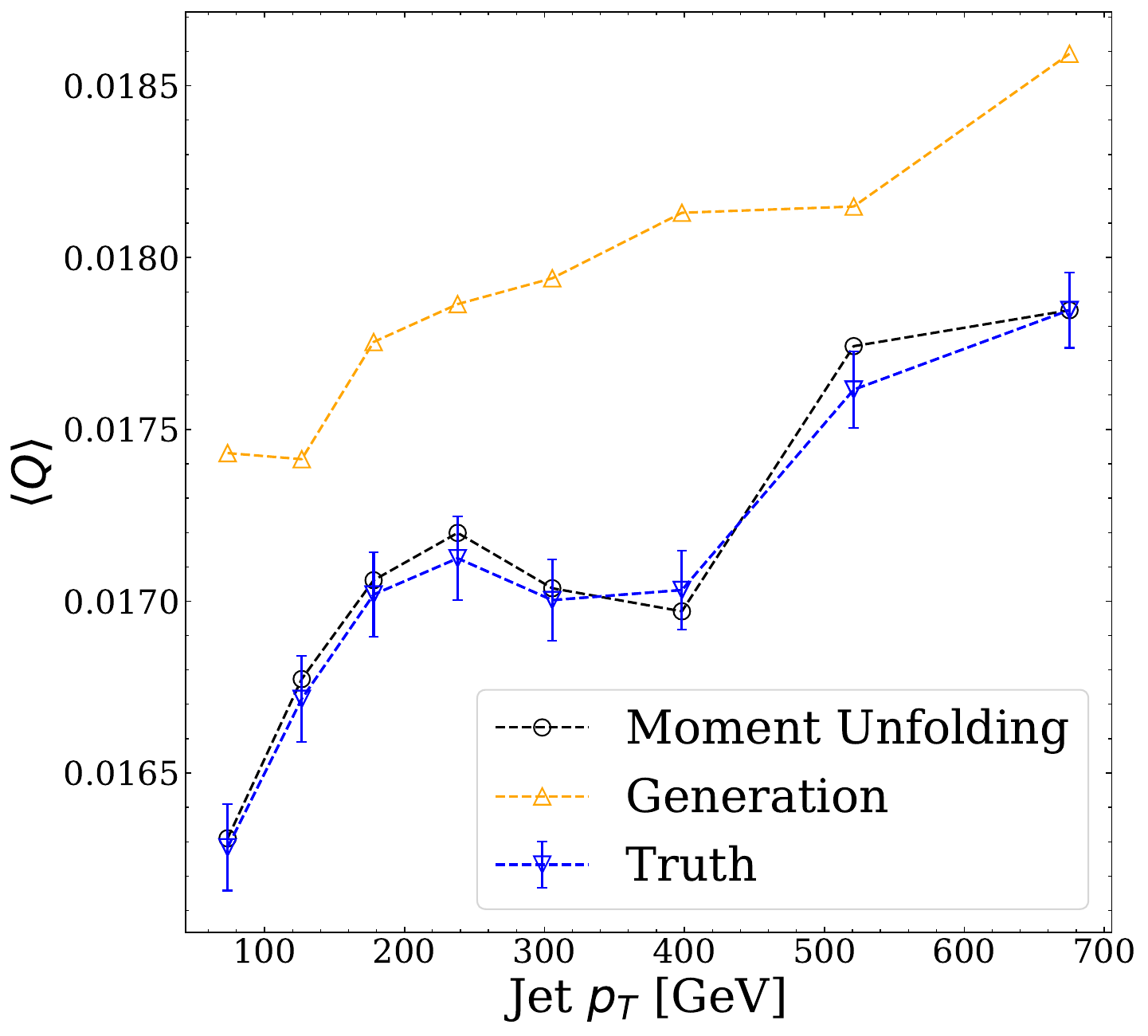}
    \label{fig:pqjetmean}}
    $\qquad\qquad$
\subfloat[]{\includegraphics[height=0.25\textwidth]{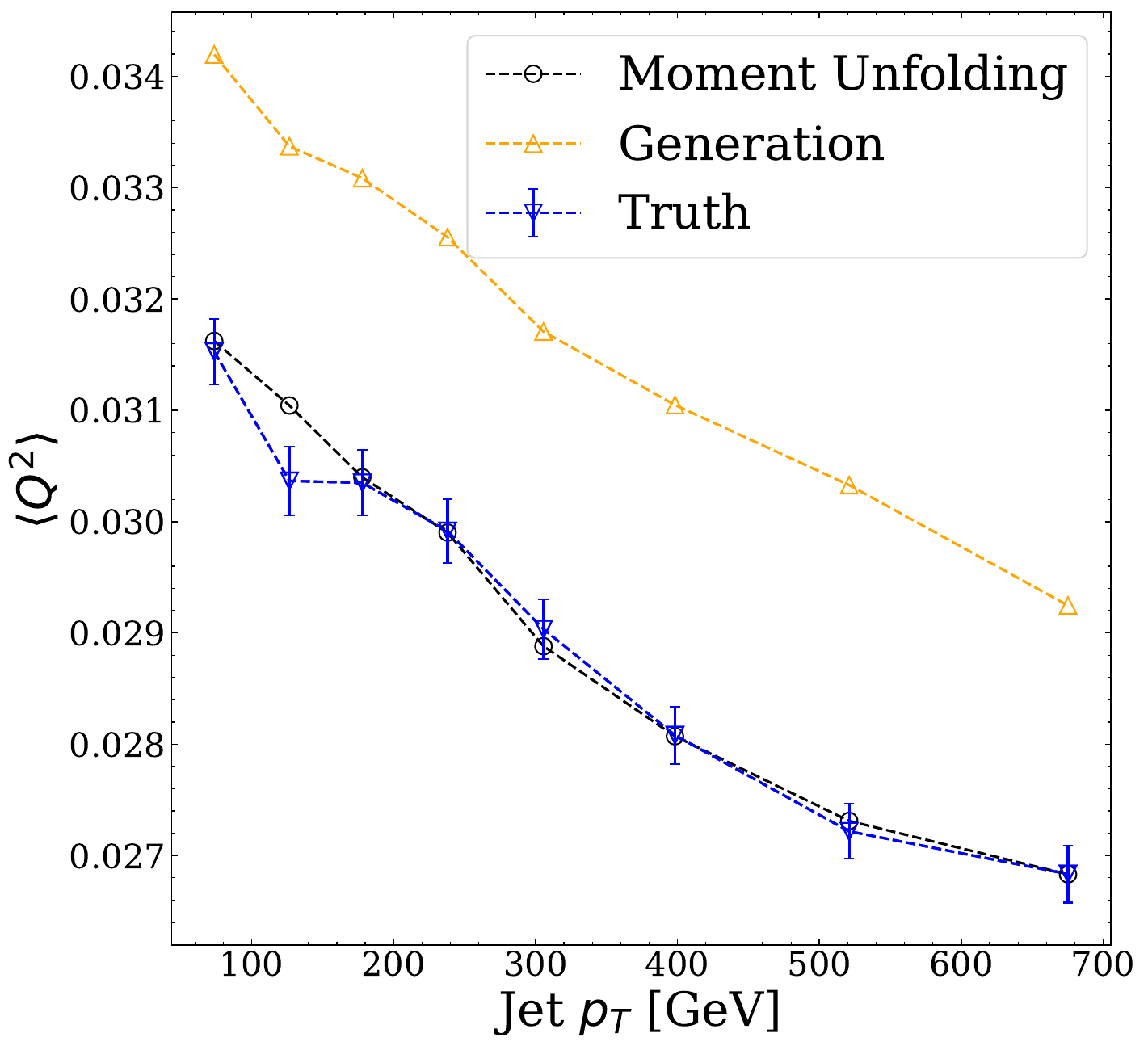}
    \label{fig:pqjetvar}}

        \subfloat[]{\includegraphics[height=0.25\textwidth]{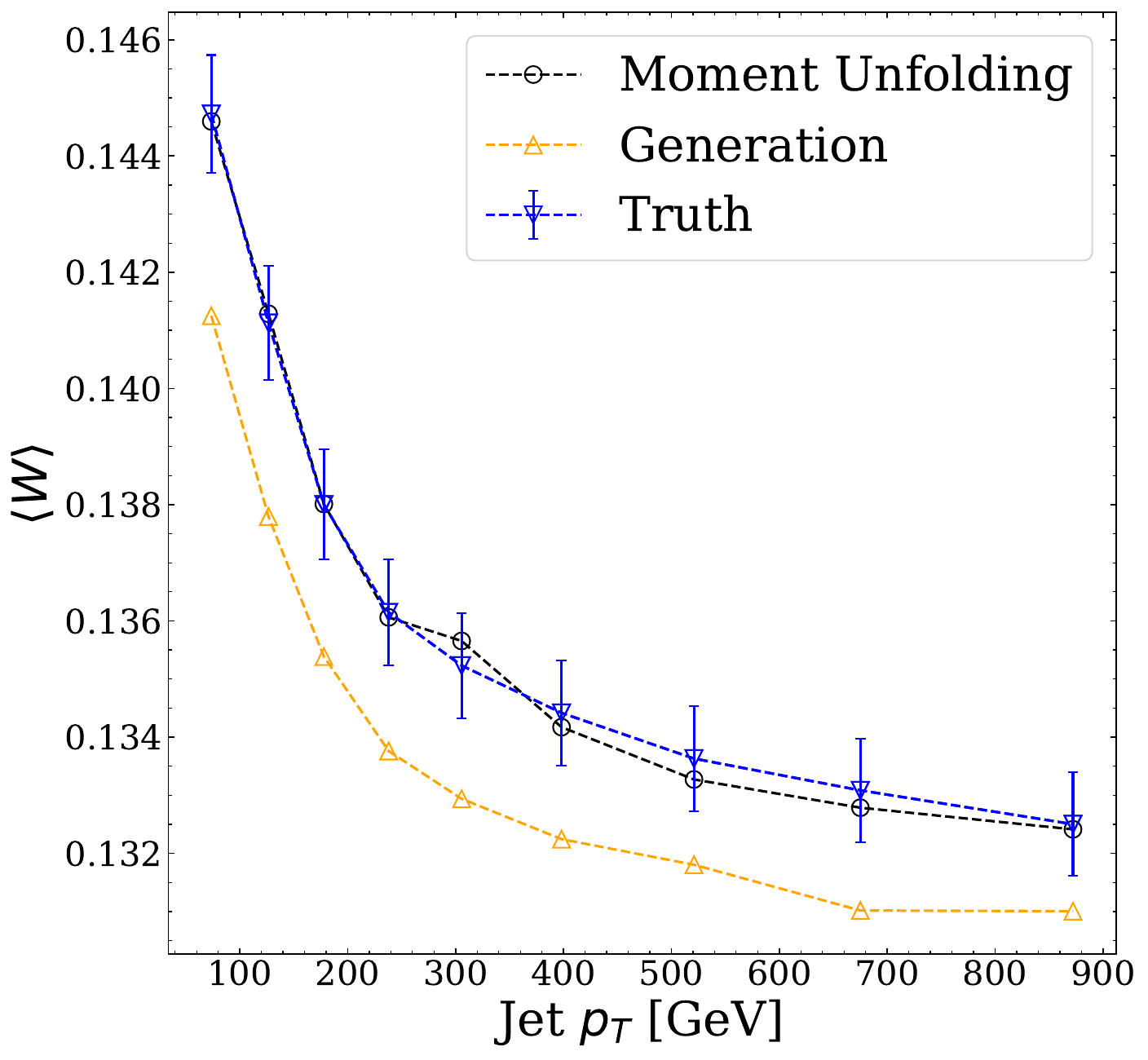}
    $\qquad\qquad$
    \label{fig:pwjetmean}}
\subfloat[]{\includegraphics[height=0.25\textwidth]{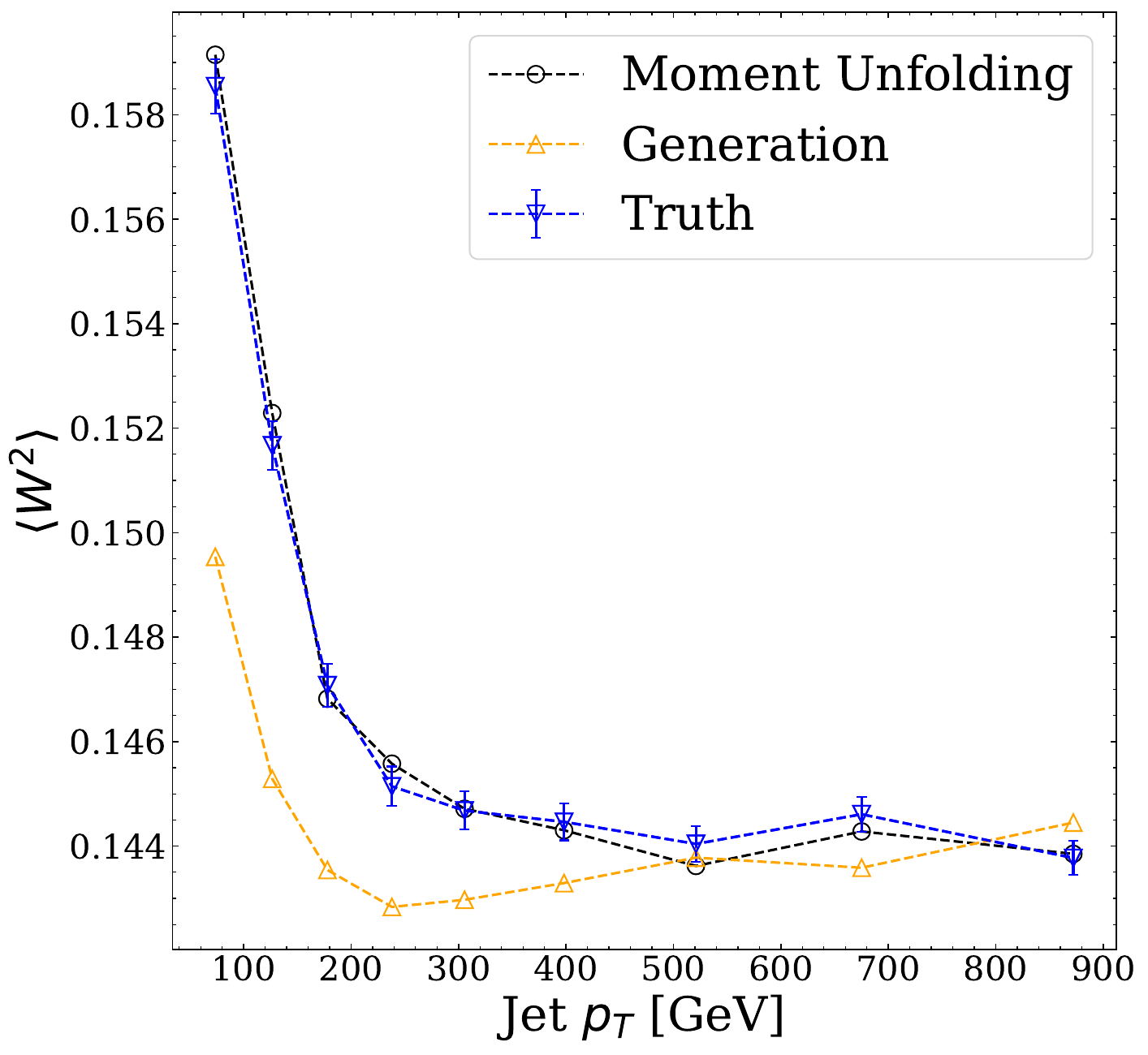}
    \label{fig:pwjetvar}}

    \subfloat[]{\includegraphics[height=0.25\textwidth]{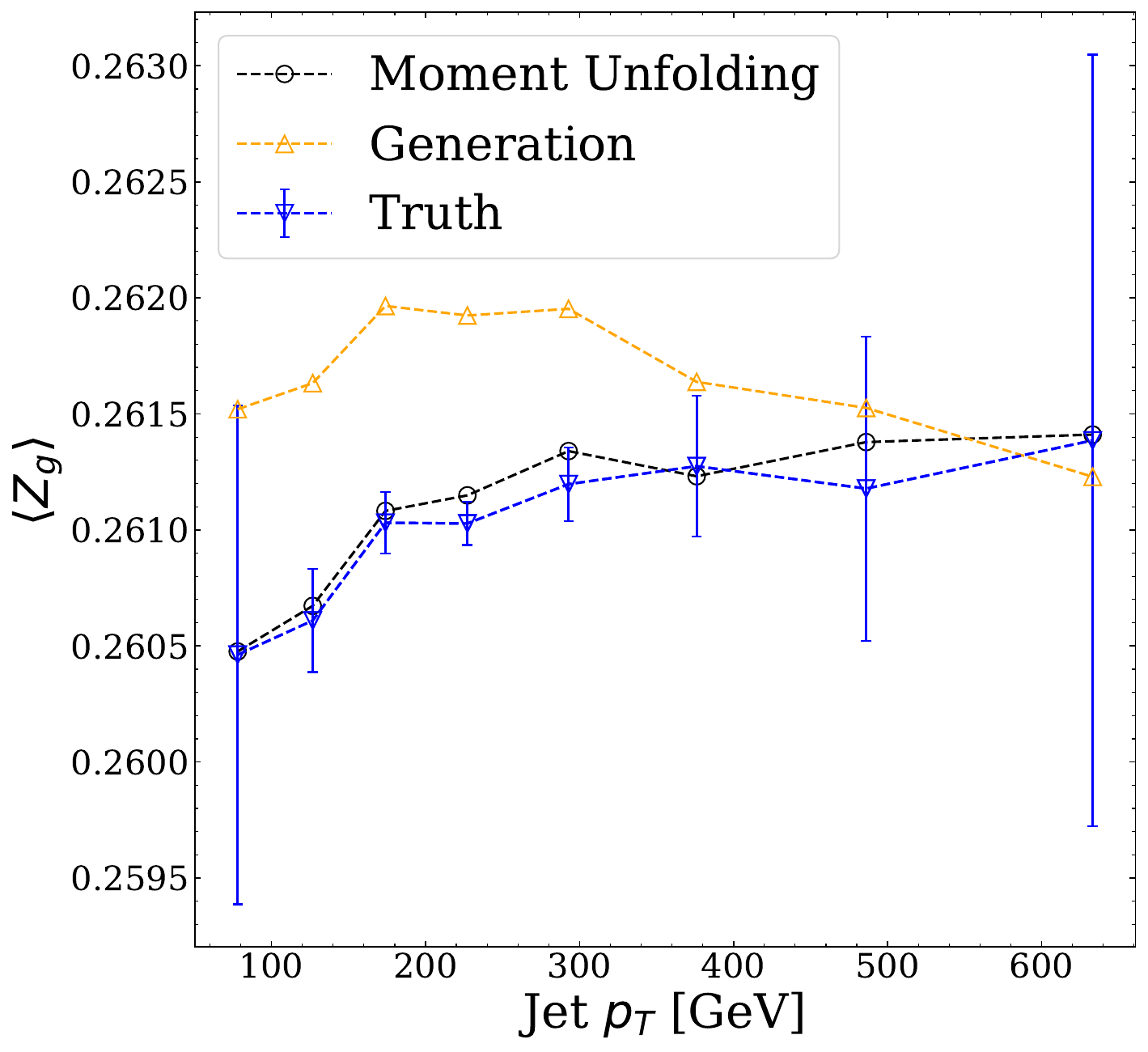}
    $\qquad\qquad$
    \label{fig:pzgjetmean}}
\subfloat[]{\includegraphics[height=0.25\textwidth]{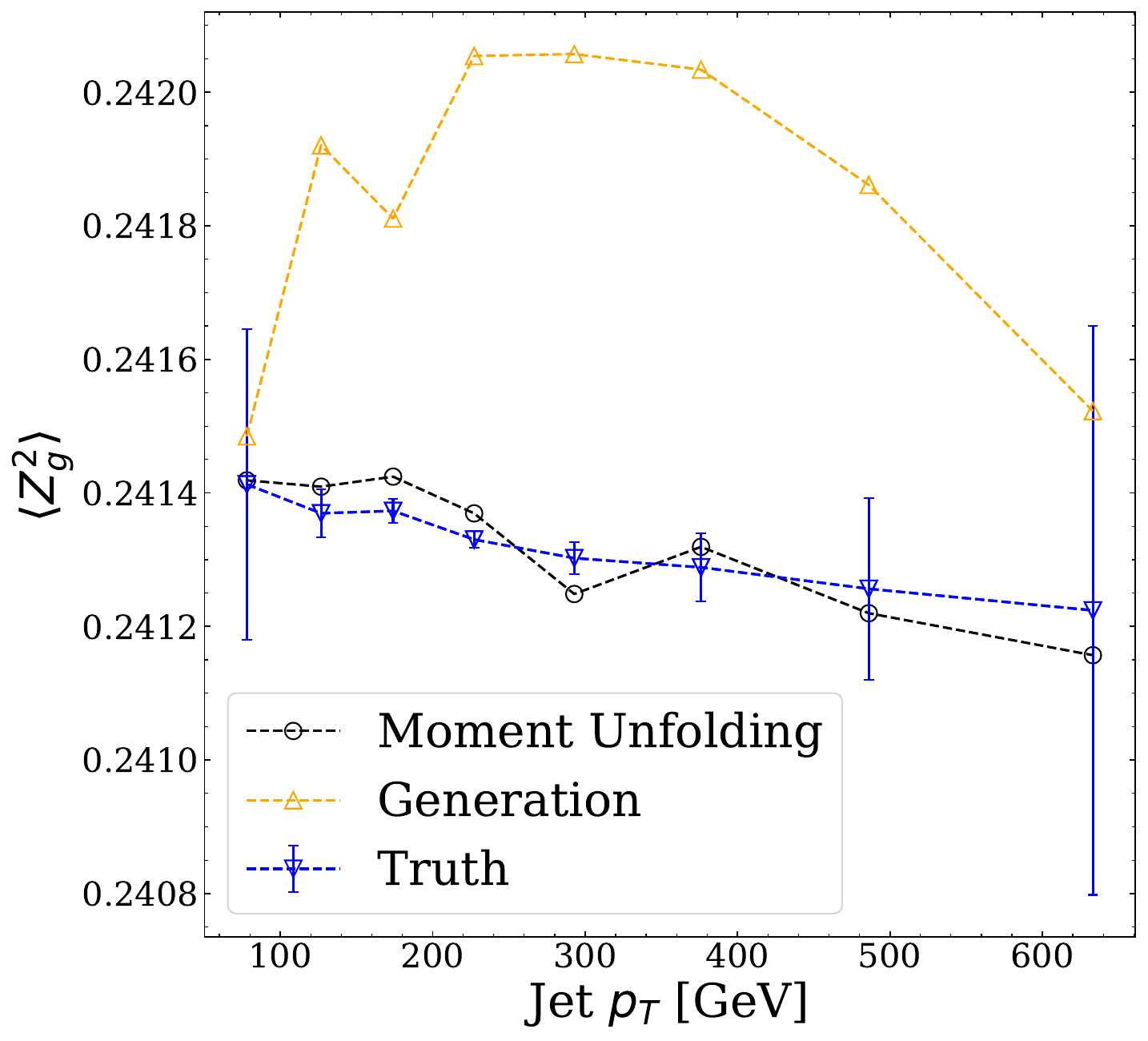}
    \label{fig:pzgjetvar}}
    
        \caption{
        The mean (left column) and variance (right column) of the jet mass (i, ii), jet charge (iii, iv), jet width (v, vi), and groomed momentum fraction (vii, viii) as a function of the transverse momentum of the jet.
        For visual clarity, only statistical uncertainties on the truth distribution are shown.
        With uncertainties, the Moment Unfolding results are in good agreement with the truth.}
    \label{fig:pjetmoments}
\end{figure*}

In \Fig{pjetmoments}, we plot the dependence of jet observable  moments on the jet $p_T$.
The left column shows the first moments of the jet mass, jet charge, and jet width, while the right column shows the corresponding second moments.
The moments are computed in bins of transverse momentum for the truth dataset (blue triangles), the generation dataset (orange triangles), and Moment Unfolding result using the weight factor $g(x; p_T)$ from \Eq{momentum_dependent_weight} (black circles).
Up to statistical uncertainties, we see that the Moment Unfolding results coincide with the moments of the truth dataset, which are substantially distinct from the moments of the generation dataset.

\subsection{Comparison to Other Methods}

Finally, we compare the unfolded moments computed through Moment Unfolding method to those obtained through three alternative unfolding methods:
\begin{itemize}
    \item \textbf{OmniFold}:  An example of unbinned unfolding;
    \item \textbf{IBU}:  An example of binned unfolding; and
    \item \textbf{IBU + Bin Correction}:  Same as above but performing the binwise correction from \Eq{bincorrection}.
\end{itemize}

\begin{figure*}
    \centering
    \subfloat[]{\includegraphics[height=0.25\textwidth]{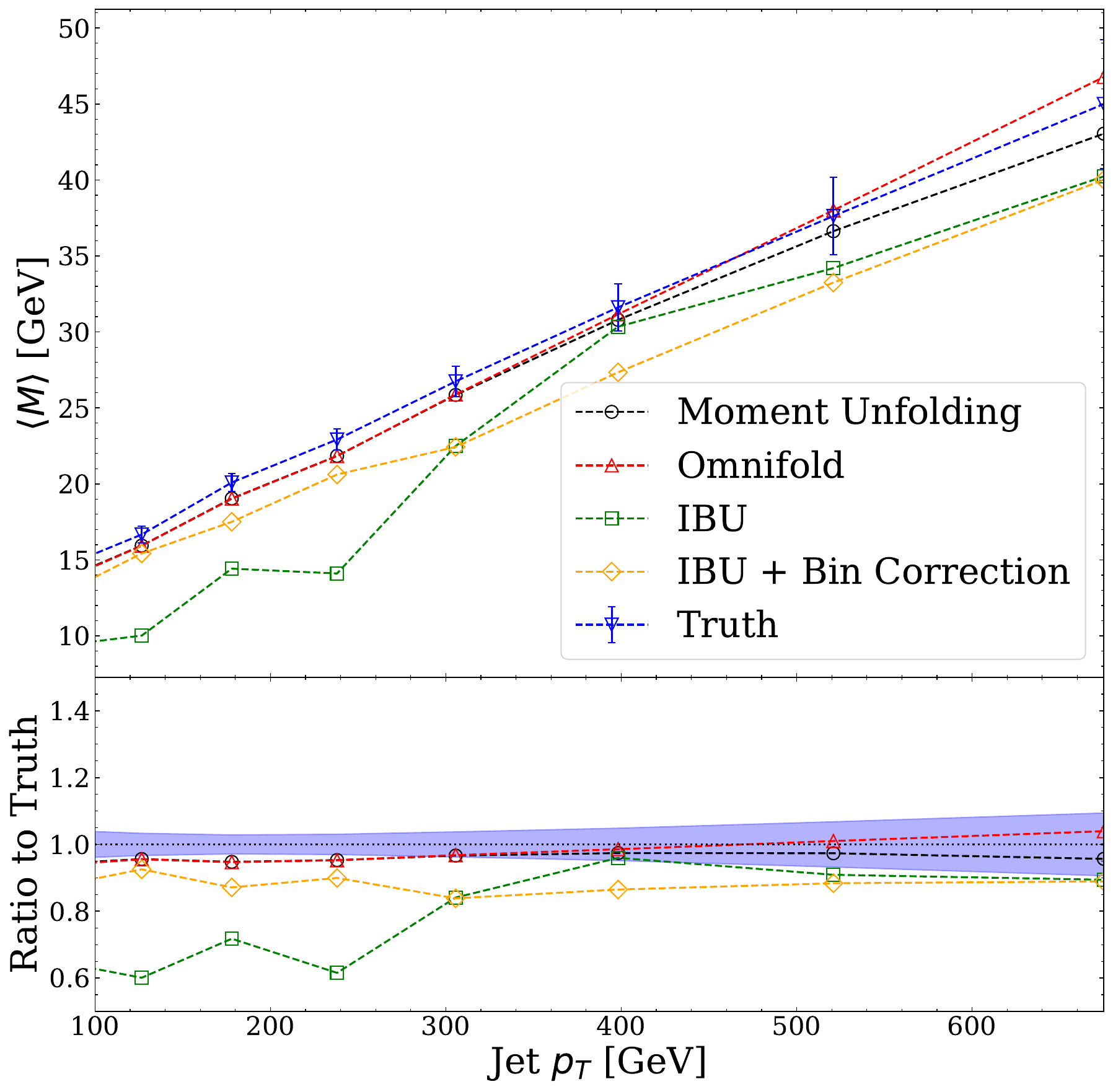}}
    $\qquad\qquad$
    \subfloat[]{\includegraphics[height=0.25\textwidth]{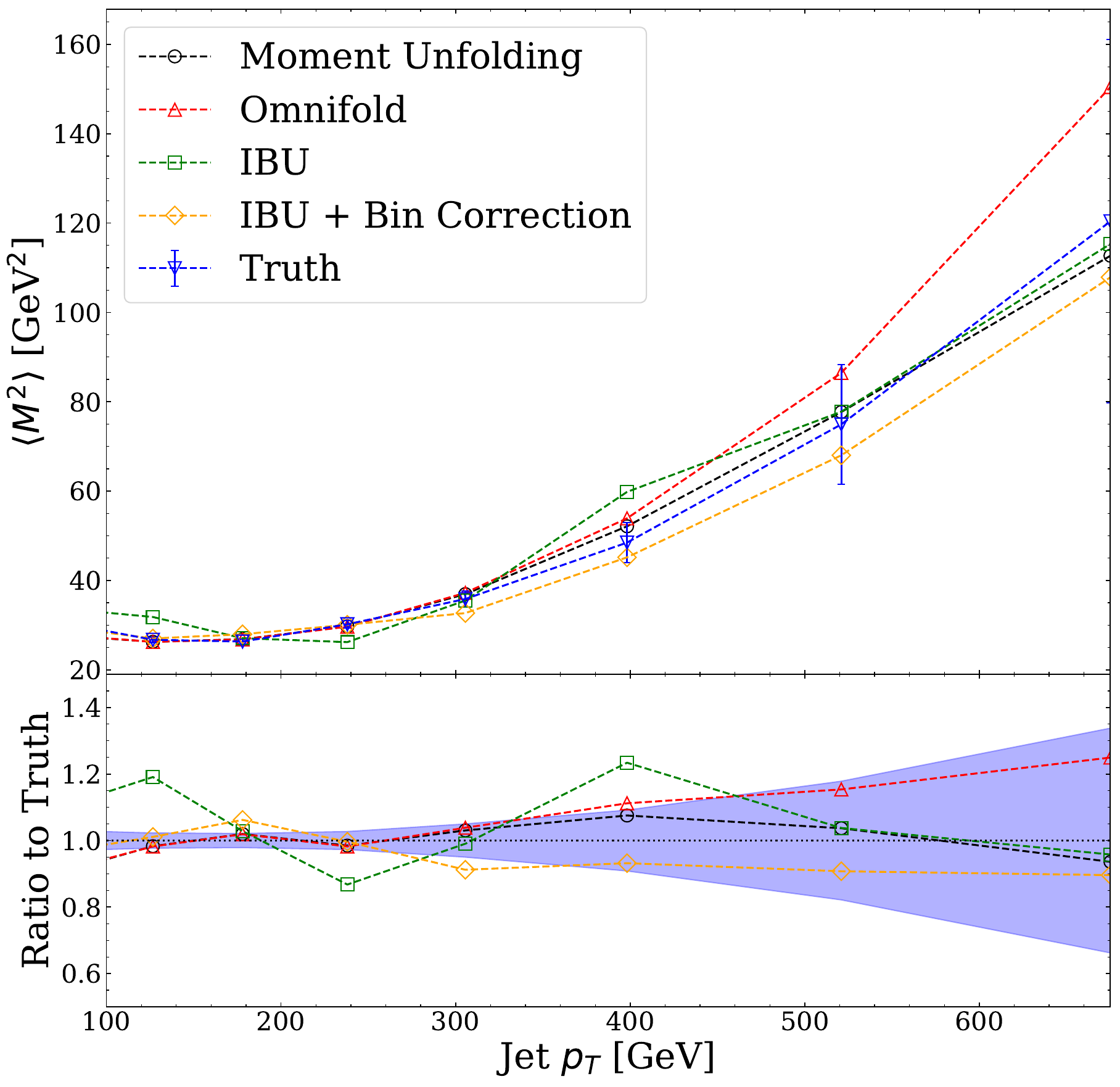}}\\

    \subfloat[]{\includegraphics[height=0.25\textwidth]{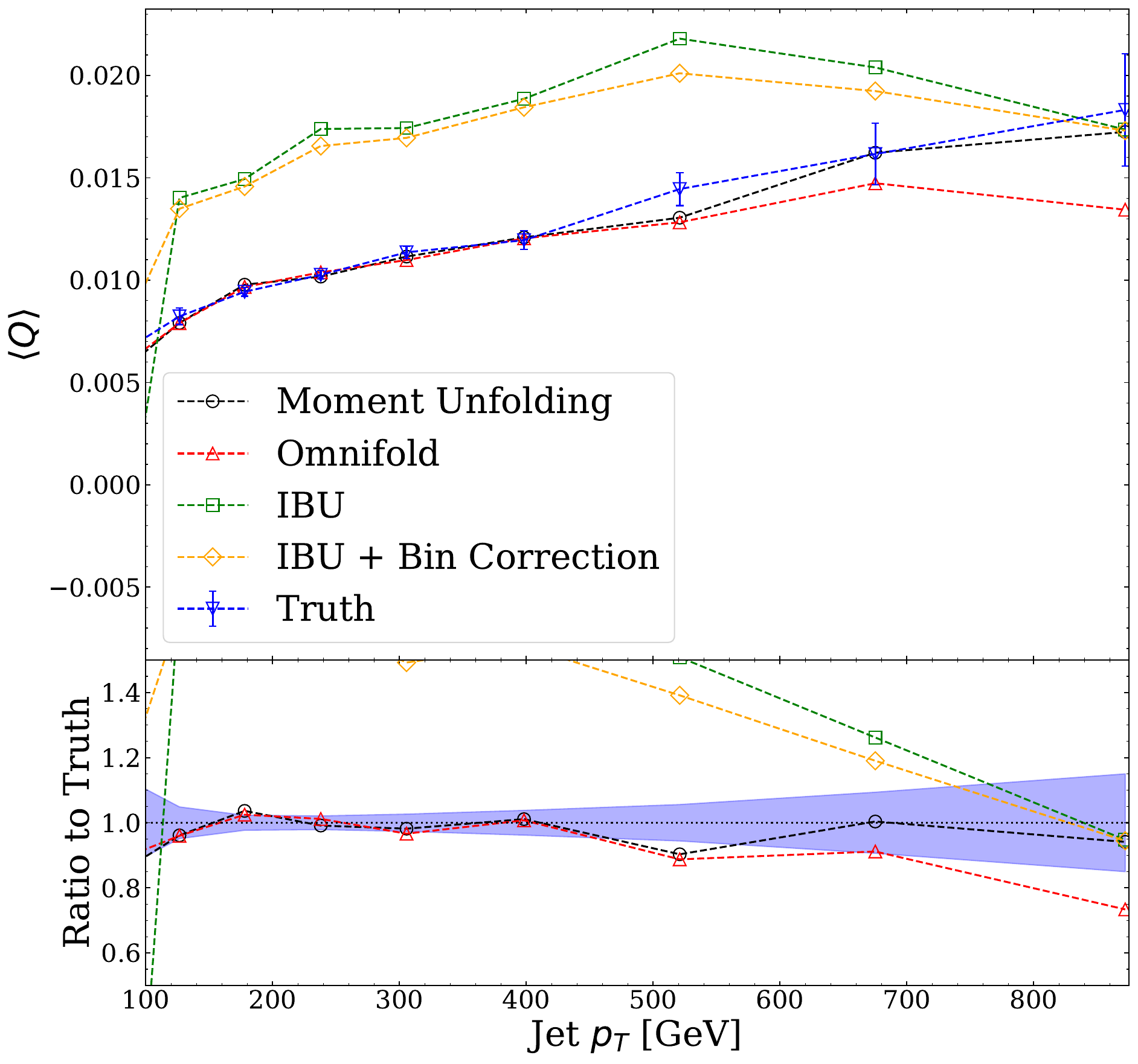}}
    $\qquad\qquad$
    \subfloat[]{\includegraphics[height=0.25\textwidth]{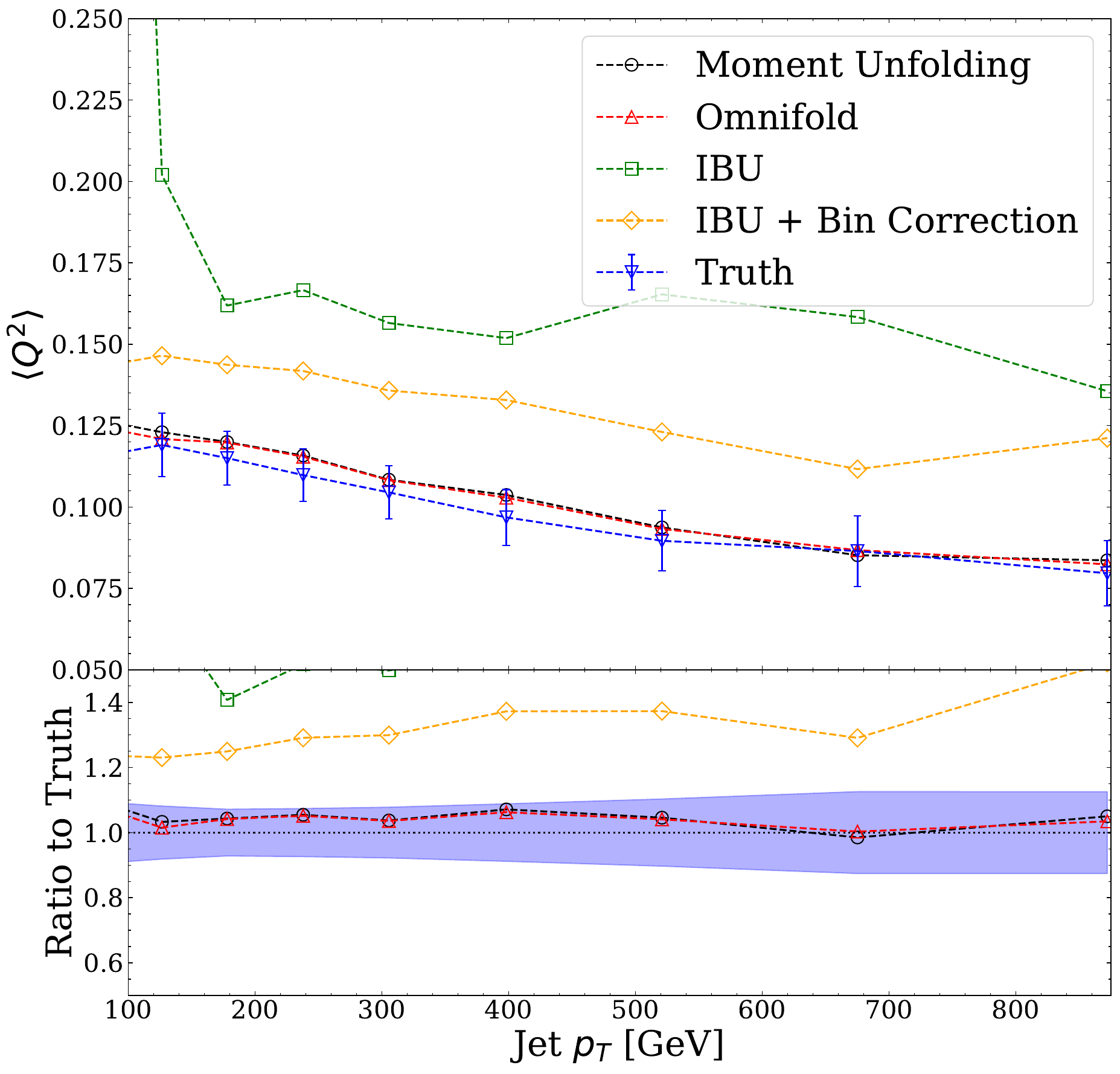}}

    \subfloat[]{\includegraphics[height=0.25\textwidth]{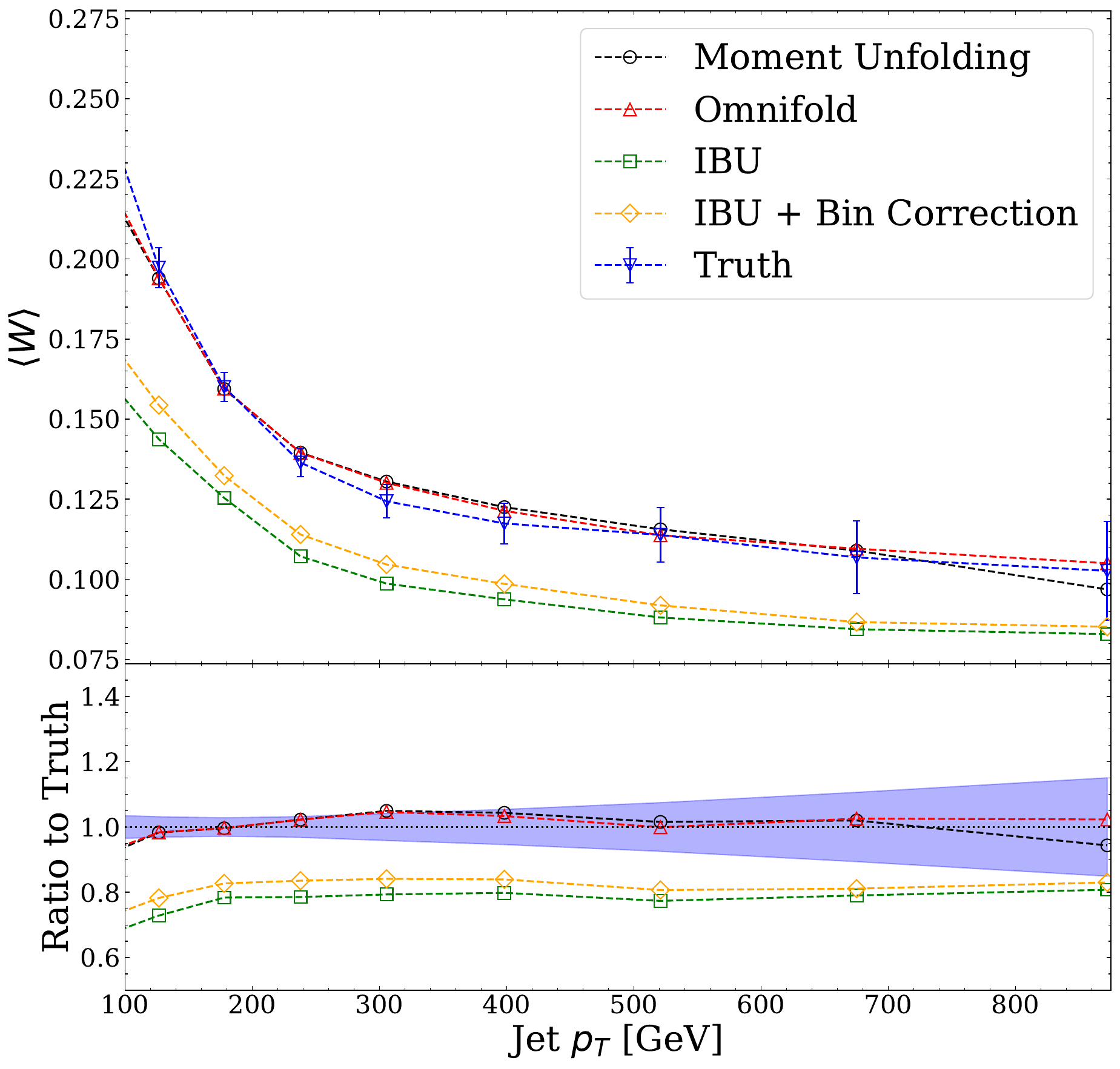}}
    $\qquad\qquad$
    \subfloat[]{\includegraphics[height=0.25\textwidth]{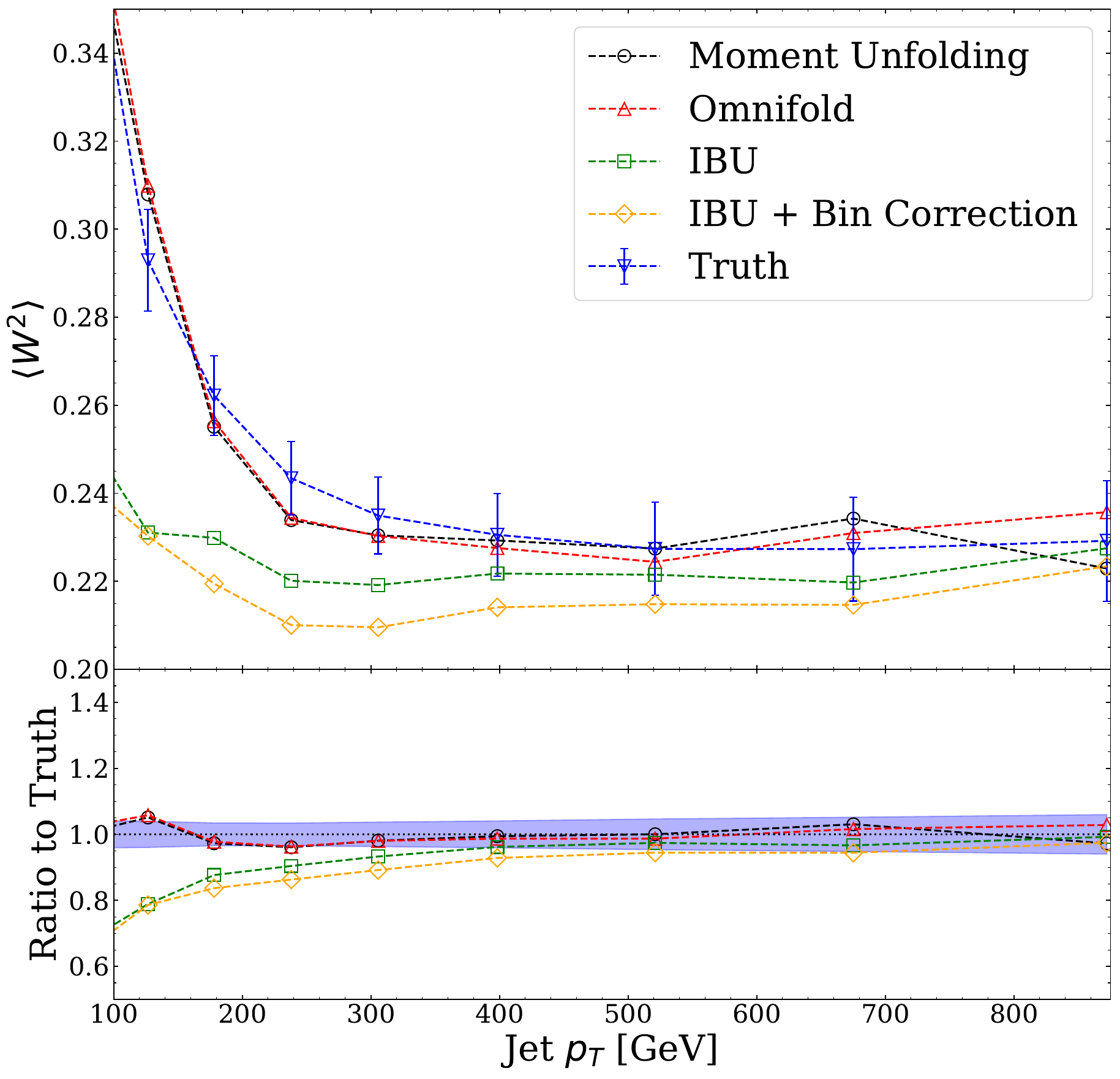}}

    \subfloat[]{\includegraphics[height=0.25\textwidth]{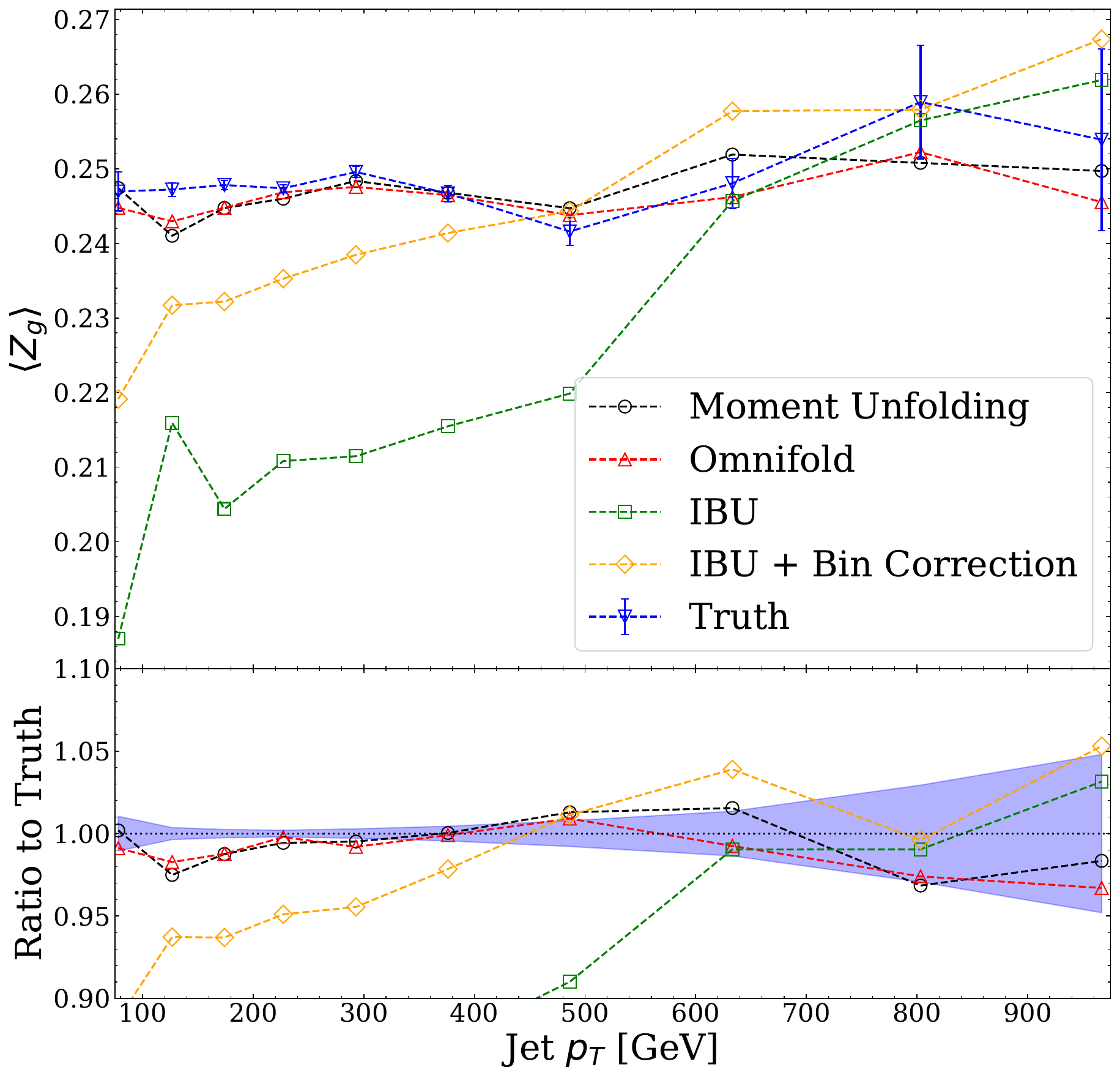}}
    $\qquad\qquad$
    \subfloat[]{\includegraphics[height=0.25\textwidth]{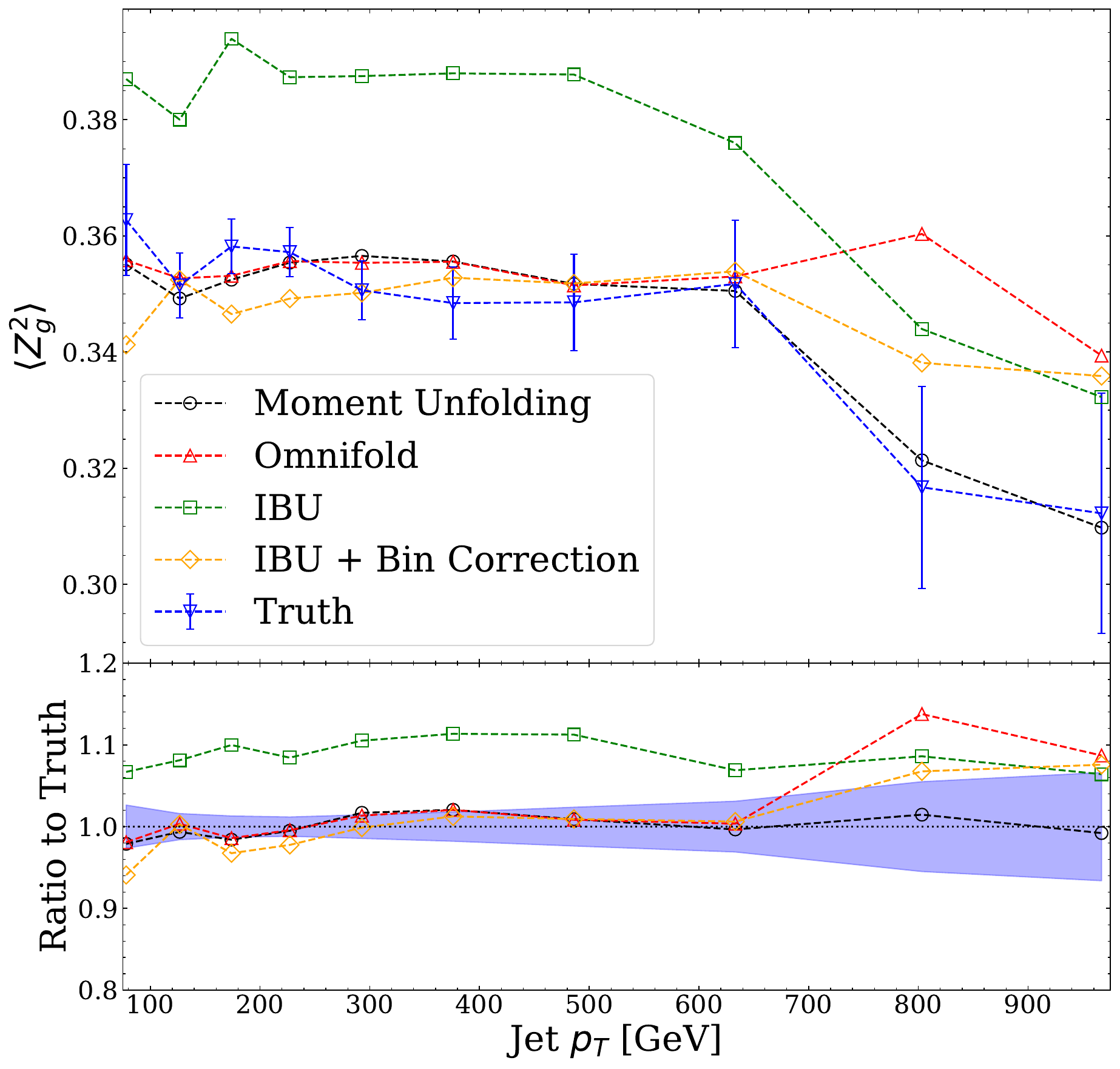}}
    
    \caption{
    Same observables and moments as \Fig{pjetmoments}, but now comparing the Moment Unfolding result to those of OmniFold, IBU, and IBU with binwise correction.
    The lower panel in each plot shows the ratio of the extracted moments to the truth as a function of jet $p_T$, demonstrating the strong performance of Moment Unfolding relative to IBU, and comparable performance to OmniFold.
    }
    \label{fig:comp}
\end{figure*}

The results of this comparison are shown in \Fig{comp}, for the first and second moments of jet mass, jet charge, jet width, and groomed momentum fraction.
The top panel of each plot shows the moments as a function of jet $p_T$, comparing Moment Unfolding (black circles), the three method listed above (IBU in green squares, IBU with the binwise correction in yellow diamonds, and Omnifold in red triangles), and the truth dataset (blue triangles).
To better highlight the performance of each method, the bottom panels of each plot show the ratio of the unfolded moments to the truth moments.
In general, the unbinned methods OmniFold and Moment Unfolding outperform both versions of IBU.
Despite having a more rigid reweighting function and simpler training paradigm, Moment Unfolding nevertheless exhibits comparable (and in some cases) better performance than OmniFold.
On average, Moment Unfolding takes about $10^4$ times longer to run than IBU, and Omnifold takes about $10^2$ times longer to run than Moment Unfolding.

\section{Conclusions and Outlook}
\label{sec:conclusions}

In this paper, we introduced the first dedicated approach to unfolding the moments of distributions without binning.
Our Moment Unfolding protocol is based on the structure of a Generative Adversarial Network, where the generator is a weighting function at particle level and the discriminator is a classifier acting at detector level.
The weight function is inspired by the Boltzmann distribution, with a small number of parameters that have a  physical interpretation as Lagrange multipliers imposing moment constraints.

Through both a simple Gaussian example and physically-relevant examples from jet physics, we showed that Moment Unfolding is able to recover the desired truth moments.
The performance is comparable to a generic approach for unbinned unfolding (OmniFold), but without the complexity of an iterative algorithm.
Moment Unfolding is able to recover moments inclusively, and with a small modification, also differential in at least one quantity.
While the dependence of one moment on one other observable is a common case, this new method can in principle be extended to more moments and differential in more quantities, though the practical challenges of this scaling is left to future studies.

Going to the extreme limit, the Moment Unfolding strategy could be extended to full distributions.
Unfolding full distributions typically requires some kind of regularization, such as limiting the number of iterations when using iterative methods.
For a non-iterative method like Moment Unfolding, one would need to regularize the functional form of the weight factor in some way, for example by using neural networks with a Lipschitz constraint~\cite{Kitouni:2021fkh,Bright-Thonney:2023gdl}.
With appropriate regularization, a generalized version of Moment Unfolding could potentially combine the flexibility of machine-learning-based approaches like OmniFold with the robustness of traditional unfolding strategies.

\section*{Code and Data}

The code for this paper can be found at \url{https://github.com/HEP-GAN/MomentUnfolding}, which makes use of \textsc{Jupyter} notebooks~\cite{Kluyver:2016aa} employing \textsc{NumPy}~\cite{harris2020array} for data manipulation and \textsc{Matplotlib}~\cite{Hunter:2007} to produce figures.
All of the machine learning was performed on an Nvidia RTX6000 Graphical Processing Unit (GPU) and running the notebook to perform unfold the first few moments of a dataset takes less than five minutes per iteration (to extract bootstrapped uncertainties we perform $500$ iterations).
The physics data sets are hosted on Zenodo at~\cite{andreassen_2019_3548091,Andreassen:2019cjw}.

\section*{Acknowledgments}

We thank Shuchin Aeron, Benjamin Fischer, and Dennis Noll for useful discussions.
BN would like to thank Stefan Kluth for discussions about unfolding moments nearly a decade ago in the context of \Reff{1509.05190}.
JT would like to thank Benoit Assi, Stefan Hoeche, and Kyle Lee for discussions of moments in the context of theory calculations.
KD and BN are supported by the U.S.\ Department of Energy (DOE), Office of Science under contract DE-AC02-05CH11231.
JT is supported by the National Science Foundation (NSF) under Cooperative Agreement PHY-2019786 (The NSF AI Institute for Artificial Intelligence and Fundamental Interactions, \url{http://iaifi.org/}), by the U.S. DOE Office of High Energy Physics under grant number DE-SC0012567, by the Simons Foundation through Investigator grant 929241, and his work was performed in part at the Aspen Center for Physics, which is supported by NSF grant PHY-2210452.

\pagebreak

\appendix

\section{Review of Boltzmann Weights}
\label{app:boltzmann}

In this appendix, we derive the weight factor from \Eq{generator}, following the familiar derivation of the Boltzmann distribution.
The goal is to learn a distribution $\ell(z)$ that optimizes the relative entropy of $\ell(z)$ with respect to a prior $q(z)$, subject to moment constraints from a distribution $p(z)$.

The KL divergence of $\ell(z)$ from $q(z)$ is:
\begin{equation}
    D_{\rm KL}(\ell \| q) = \int \ell(z) \log \frac{\ell(z)}{q(z)} \, \dd z.
\end{equation}
In the absence of constraints, this quantity would be minimized when $\ell(z) = q(z)$.
Note that the KL divergence is not symmetric between $\ell(z)$ and $q(z)$, which is essential to the following derivation.

To impose the moment constraints, we include Lagrange multipliers $\beta_a$ to force $\ell(z)$ to have the same first $n$ moments as $p(z)$.
This corresponds to the loss function:
\begin{equation}
L = D_{\rm KL}(\ell \| q) + \sum_{a = 0}^n \beta_a \int z^a \big(\ell(z) - p(z)\big) \, \dd z,
\end{equation}
where the $a=0$ term enforces that $\ell(z)$ is properly normalized.
Taking a functional derivative of the loss with respect to $\ell(z)$:
\begin{equation}
    \frac{\delta L}{\delta \ell(z)} = \log \frac{\ell(z)}{q(z)} + 1 + \sum_{a = 0}^n \beta_a z^a.
\end{equation}
Setting this to zero to find the minimum, the solution is:
\begin{equation}
\ell(z)  = q(z) \exp\Bigg[- 1 - \sum_{a = 0}^n \beta_a z^a \Bigg].
\end{equation}

Writing $\ell(z) = g(z) \, q(z)$ and solving $\beta_0$ for the normalization condition, we recover the desired weight factor from \Eq{generator}, repeated for convenience:
\begin{equation}
        g(z) = \frac1{P}\exp\Bigg[-\sum_{a = 1}^n\beta_a\,z^a\Bigg]. 
\end{equation}
Here, the normalization factor is:
\begin{equation}
    P = \int q(z) \exp\Bigg[-\sum_{a = 1}^n\beta_a\,z^a\Bigg] \, \dd z.
\end{equation}
The remaining Lagrange multipliers $\beta_a$ are determined by solving the moment constraints, which do not have a closed form in general.

\section{Analytic Closure Tests}
\label{app:analytic}

In this appendix, we derive the asymptotic conditions under which Moment Unfolding will correctly recover the moments of the true distribution.

\subsection{Perfect Detector Response}

We start with the case of perfect detector response, such that we can work entirely with particle-level distributions in $z$.
The derivation in \App{boltzmann} shows that the weight factor from \Eq{generator} optimizes the relative entropy subject to the moment constraints.
Here, we prove that minimizing the MLC loss with respect to the weight factor parameters recovers the desired moments.
This is a non-trivial closure test of Moment Unfolding, since loss functions other than MLC do not generically satisfy this property.

Let $p(z)$ be the true particle-level distribution, $q(z)$ be the particle-level generator, and $g(z)$ be the weight factor from \Eq{generator}.
For convenience, we define the reweighted distribution after Moment Unfolding as:
\begin{equation}
    \qt(x) = q(z) \, g(z).
\end{equation}
Asymptotically, the MLC loss of $p(z)$ to $\qt(z)$ is:
\begin{equation}
    L = - \int p(z) \log d(z) \, \dd z - \int \qt(z) \big(1-d(z)\big) \, \dd z,
\end{equation}
where $d(z)$ is the discriminator function.
Because we are assuming perfect detector response, it makes sense to talk about the discriminator acting on particle-level quantities.

Taking a functional derivative of the loss with respect to $d(z)$, we find:
\begin{equation}
    \frac{\delta L}{\delta d(z)} = - \frac{p(z)}{d(z)} + \qt(z).
\end{equation}
Setting this equal to zero, the optimal discriminator is $d_*(z) = p(z) / \qt(z)$.
Plugging this back into the MLC loss and using the fact that $p(z)$ and $\qt(z)$ are both normalized, we find:
\begin{equation}
\label{eq:loss_with_opt_discriminator}
    L \big|_{d = d_*} = - \int p(z) \log \frac{p(z)}{\qt(z)} \dd z = - D_{\rm KL}(p \| \qt).
\end{equation}
Note that this result is not symmetric between $p(z)$ and $\qt(z)$, which is essential to the rest of the derivation.

We now want to optimize the generator parameters assuming the optimal discriminator.
Letting $\widetilde{L} \equiv L \big|_{d = d_*}$ for notational convenience, the derivative of the discriminator-optimized loss with respect to the generator parameters is:
\begin{equation}
\label{eq:loss_vary_beta_a}
    \frac{\partial \widetilde{L}}{\partial \beta_a} = \int \frac{\delta \widetilde{L}}{\delta g(z)} \frac{\partial g(z)}{\partial \beta_a} \, \dd z.
\end{equation}
The functional derivative of the loss with respect to the generator is:
\begin{equation}
    \frac{\delta \widetilde{L}}{\delta g(z)} = \frac{p(z)}{g(z)},
\end{equation}
while the derivatives of the generator with respect to its parameters are:
\begin{equation}
\label{eq:generator_derivative_wrt_parameters}
    \frac{\partial g(z)}{\partial \beta_a} = -g(z) \Big(z^a + \frac{1}{P}\frac{\partial P}{\partial \beta_a} \Big) = -g(z) \big(z^a - \ev{Z^a}_\qt \big).
\end{equation}
Therefore, setting \Eq{loss_vary_beta_a} equal to zero is equivalent to enforcing:
\begin{equation}
\label{eq:perfect_detector_moment_relation}
    \ev{Z^a}_\qt = \ev{Z^a}_p,
\end{equation}
for all $a = 1, \ldots ,n$.
This proves that the learned moments match those from data, at least in the asymptotic limit assuming perfect detector response and optimal learning.

\subsection{Universal Detector Response}

In the case of a realistic detector, the learned moments will not in general match the truth moments.
That said, the deviations from closure will be small as long as detector distortions are small.

Crucially, all unfolding methods assume that the detector response is universal between real data and simulation.
This means that the detector-level distributions can be written in terms of a universal response function $r(x|z)$ as:
\begin{align}
    p(x) &= \int r(x|z) \, p(z) \, \dd z,\\
    q(x) &= \int r(x|z) \, q(z) \, \dd z.
\end{align}
The reweighted distribution after Moment Unfolding is
\begin{equation}
    \qt(x) = \int r(x|z) \, q(z) \, g(z) \, \dd z.
\end{equation}
In the case that $r(x|z) = \delta(x-z)$, we recover the closure result from \Eq{perfect_detector_moment_relation}.
We now derive the moment relation in the case that the detector is imperfect but still universal.

The discriminator $d(x)$ now acts on detector-level quantities.
Apart from swapping $z$ for $x$, though, the derivation of the discriminator-optimized MLC loss is the same as in \Eq{loss_with_opt_discriminator}:
\begin{equation}
    \widetilde{L} = - \int p(x) \log \frac{p(x)}{\qt(x)} \, \dd x.
\end{equation}
The derivative of this with respect to the generator parameters is somewhat more involved than \Eq{loss_vary_beta_a}:
\begin{equation}
\label{eq:loss_vary_beta_a_universal}
    \frac{\partial \widetilde{L}}{\partial \beta_a} = \int \frac{\delta \widetilde{L}}{\delta \qt(x)} \frac{\delta \qt(x)}{\delta g(z)} \frac{\partial g(z)}{\partial \beta_a} \, \dd z \, \dd x.
\end{equation}
The functional derivatives are:
\begin{align}
    \frac{\delta \widetilde{L}}{\delta \qt(x)} &= \frac{p(x)}{\qt(x)}, \\
    \frac{\delta \qt(x)}{\delta g(z)} & = r(x|z) \, q(z).
\end{align}
The $\partial g(z) / \partial \beta_a$ derivative is the same as \Eq{generator_derivative_wrt_parameters}.

Setting \Eq{loss_vary_beta_a_universal} equal to zero, we find that the learned moments satisfy:
\begin{equation}
\label{eq:universal_detector_moment_relation}
    \ev{Z^a}_\qt = \ev{Z^a}_{p_{\rm mod}},
\end{equation}
where the modified data distribution is:
\begin{equation}
\label{eq:p_mod_def}
    p_{\rm mod}(z) = \int \frac{p(x) \, r(x | z)\, \qt(z)}{\qt(x)} \, \dd x.
\end{equation}
Thus, the moments of $\qt(z)$ match the moments of $p_{\rm mod}(z)$, which are in general different from those of $p(z)$.

To better interpret $p_{\rm mod}(z)$, it is convenient to rewrite it in the following form:
\begin{equation}
p_{\rm mod}(z) = \int f(z | z') \, p(z') \, \dd z',
\end{equation}
where $f(z | z')$ can be thought of as a transfer function that maps the actual particle-level truth information to the learned particle-level information.
In the case of perfect detector response, $f(z | z') = \delta(z-z')$.
For a realistic detector, one can manipulate \Eq{p_mod_def} to find:
\begin{equation}
    f(z|z') = \int \qt(z | x ) \, r(x | z') \, \dd x.
\end{equation}
Here, $\qt(z|x) = r(x | z) \, \qt(z) / \qt(x)$ is the \emph{inverse} detector response derived from the reweighted simulation, which is in general not universal.

Thus, one can interpret $f(z | z')$ as taking the particle-level information, passing it through the universal detector response, and pulling it back through the non-universal reweighted simulation.
To the extent that the reweighted simulation is sufficiently similar to the real data, $p_{\rm mod}(z)$ will be similar enough to $p(z)$ to satisfy closure.
One obstruction to closure is if the detector response includes large distortions, such that the inverse detector response is highly dependent on the reweighting function.
Another obstruction is if the particle-level generator has poor overlapping support with the truth, such that the reweighting function needs to be large in poorly modeled regions of phase space.
Both of these obstructions are common to all unfolding methods, though, and not unique to Moment Unfolding.

\pagebreak

\bibliography{HEPML, main}

\end{document}